\documentclass[11pt,numbers,letter,preprint,sort&compress]{aastex}
\bibpunct{(}{)}{,}{n}{,}{,}
\usepackage{rotating}
\usepackage{float}
\usepackage{longtable}
\usepackage{graphicx}
\usepackage{amssymb}
\usepackage[cm]{fullpage}
\begin{document}
\title{Update on the correlation of the highest energy cosmic rays\\ with nearby extragalactic matter}
\author{
\par\noindent
{\bf The Pierre Auger Collaboration} \\
P.~Abreu$^{73}$, 
M.~Aglietta$^{55}$, 
E.J.~Ahn$^{89}$, 
D.~Allard$^{31}$, 
I.~Allekotte$^{1}$, 
J.~Allen$^{92}$, 
J.~Alvarez Castillo$^{66}$, 
J.~Alvarez-Mu\~{n}iz$^{80}$, 
M.~Ambrosio$^{48}$, 
A.~Aminaei$^{67}$
% ...

\small
\par\noindent
$^{1}$ Centro At\'{o}mico Bariloche and Instituto Balseiro (CNEA-
UNCuyo-CONICET), San Carlos de Bariloche, Argentina \\
% ...
$^{31}$ Laboratoire AstroParticule et Cosmologie (APC), 
Universit\'{e} Paris 7, CNRS-IN2P3, Paris, France \\
% ...
$^{48}$ Universit\`{a} di Napoli "Federico II" and Sezione INFN, 
Napoli, Italy \\
% ...
$^{55}$ Istituto di Fisica dello Spazio Interplanetario (INAF),
 Universit\`{a} di Torino and Sezione INFN, Torino, Italy \\
% ...
$^{66}$ Universidad Nacional Autonoma de Mexico, Mexico, D.F., 
Mexico \\
$^{67}$ IMAPP, Radboud University, Nijmegen, Netherlands \\
% ...
$^{73}$ LIP and Instituto Superior T\'{e}cnico, Lisboa, Portugal \\
% ...
$^{80}$ Universidad de Santiago de Compostela, Spain \\
% ...
$^{89}$ Fermilab, Batavia, IL, USA \\
% ...
$^{92}$ New York University, New York, NY, USA \\
% ...
\par\noindent
($\dagger$) Deceased \\
(a) at Konan University, Kobe, Japan \\
(c) at Caltech, Pasadena, USA \\
}
\newpage
\begin{abstract}
Data collected by the Pierre Auger Observatory through 31 August 2007
showed evidence for anisotropy in the arrival directions of cosmic
rays above the Greisen-Zatsepin-Kuz'min energy threshold, \nobreak{$6\times 10^{19}$~eV}.
The anisotropy was measured by the fraction of arrival directions that
are less than $3.1^\circ$ from the position of an active galactic nucleus 
within 75 Mpc (using the V\'eron-Cetty and V\'eron $12^{\rm th}$ catalog).  
An updated measurement of this fraction is reported here using the arrival directions 
of cosmic rays recorded above the same energy threshold through 31 December 2009.
The number of arrival directions has increased from 27 to 69,
allowing a more precise measurement. The correlating fraction is
$(38^{+7}_{-6})\%$, compared with $21\%$ expected for isotropic cosmic
rays.  This is down from the early estimate of $(69^{+11}_{-13})\%$.
The enlarged set of arrival directions is examined also in relation to
other populations of nearby extragalactic objects: galaxies in the 2
Microns All Sky Survey and active galactic nuclei detected in hard X-rays by the
Swift Burst Alert Telescope.  A celestial region around the position
of the radiogalaxy Cen A has the largest excess of arrival directions
relative to isotropic expectations.  The 2-point autocorrelation
function is shown for the enlarged set of arrival directions and
compared to the isotropic expectation.
\end{abstract}
\maketitle

\section{Introduction}

The astrophysical sites of origin of ultra high-energy cosmic rays
(UHECRs) remain elusive after almost a half century since a cosmic ray
(CR) with energy around $10^{20}$~eV was first reported \cite{linsley}. 
Anisotropy in the arrival directions of UHECRs is expected to
provide significant clues for identifying their sources. Protons and
nuclei with these energies interact with the cosmic microwave
background (CMB), either by pion photoproduction or by nuclear
photodisintegration. This interaction limits the distance from which a
source can contribute significantly to the flux on Earth, as predicted
by Greisen~\cite{g} and by Zatsepin and Kuz'min~\cite{zk} (the GZK effect). 
For instance, most of the observed flux above 60~EeV (1~EeV~$\equiv10^{18}$~eV) should come from sources within a ``GZK horizon'' which
is approximately 200~Mpc. Processes that could accelerate particles up
to such energies require special astrophysical conditions
\cite{hillasconditions}. Few classes of astrophysical objects, such as
active galactic nuclei, radio-galaxy lobes and sources of gamma-ray bursts, 
meet these requirements. Inhomogeneities in their spatial distribution
within the GZK horizon may imprint a detectable anisotropy in the
UHECR arrival directions. Comparing the arrival directions with the
celestial positions of different types of astronomical objects is a
useful tool for identifying the sources provided intervening magnetic
fields do not deflect the cosmic ray trajectories through large
angles.

The flux of UHECRs is extraordinarily small, approximately one
particle per square kilometre per century above 60~EeV.  Large
detection areas are essential. This is achieved by measuring the
cosmic rays indirectly through the extensive air showers (EAS) that
they produce in the atmosphere. Two complementary techniques are
currently used: the measurement of the fluorescence light induced in
the atmosphere by the particles in the EAS and the detection of the
secondary particles at ground level using an array of surface
detectors.  The Pierre Auger Observatory implements air fluorescence
and water-Cherenkov detection in a hybrid instrument with an aperture
of 7000~km$^2$sr.  The implementation of the baseline design for the 
Southern Auger Observatory in Argentina \cite{paoproperties} was 
completed in June 2008.

Using data collected through 31 August 2007, the Pierre Auger
Collaboration reported in \cite{pao1,pao2} a correlation between the
arrival directions of UHECRs with energies exceeding 56 EeV and the
positions of nearby objects from the $12^{\rm th}$ edition of the
catalog of quasars and active galactic nuclei (AGNs) by V\'{e}ron-Cetty
and V\'{e}ron \cite{vcv} (VCV catalog). The null hypothesis of
isotropy was rejected with 99\% confidence based on a single-trial
test that was motivated by early data and confirmed by data collected 
subsequent to the definition of the test.  This correlation with 
nearby extragalactic objects is consistent with cosmic rays from more
distant sources having lost energy in accordance with the flux
suppression seen in the measured energy spectrum
~\cite{paoflux,paoflux2,hiresflux} and the GZK expectation.  However, the VCV
correlation is not sufficient to identify individual sources or a
specific class of astrophysical sites of origin. The VCV catalog is a
compilation of known AGNs that is neither homogeneous nor
statistically complete. Moreover, active galaxies in this catalog trace
the nearby large scale matter distribution, and that 
includes all types of candidate astrophysical sources, not only AGNs
and their subclasses.  Analyses comparing the Auger data reported in
\cite{pao1,pao2} with different types of nearby extragalactic objects
can be found in 
\cite{kashti,george,ghisellini,bariloche,nagar,hillas,moskalenko,
glennys,wibig,takami,tinyakov}.

This paper reports the arrival directions of CRs measured
with the Pierre Auger Observatory up to 31 December 2009 that have
energies above the same threshold as those reported in
\cite{pao1,pao2}. The data set has increased from 27 to 69 CR events,
and is described in section~\ref{sec:dataset}.

In section~\ref{vcv} we update the measured fraction of CR 
arrival directions which correlate with the positions of objects in the VCV
catalog. The measurement uses identical parameters as in
the test reported in \cite{pao1,pao2}.  

In section~\ref{othercatalogs} we examine the 69 arrival directions with
regard to their correlation with populations of nearby extragalactic
objects characterised by alternative catalogs. We compare the pattern
of the arrival directions with that of the overall matter distribution
in the local universe as traced by the galaxies in the 2MASS Redshift
Survey (2MRS)  \cite{2MRS,2MASS}, which is the most densely sampled 
all-sky redshift survey to date, 
and with AGNs detected in X-rays with the Swift Burst Alert
Telescope (BAT) \cite{SWIFT22,SWIFT58}. 

In section~\ref{other} the intrinsic clustering properties of arrival
directions are characterised using their autocorrelation function.
We also analyse the region with the largest excess of
arrival directions compared to isotropic expectations. 

We summarise the results and potential
implications in section~\ref{conclusions}.  
Some details relating to the 69 UHECRs above 55 EeV 
are tabulated in the appendix.\footnote{The list of the first 27 events 
was published in \cite{pao2}. Since then, the reconstruction algorithms and
  calibration procedures of the Pierre Auger Observatory have been
  updated and refined. The lowest energy among the same 27 events
  (which was 57~EeV in \cite{pao2}) is 55~EeV according to the latest
  reconstruction.}

\section{The Observatory and the dataset}
\label{sec:dataset}

The Pierre Auger Southern Observatory is located in the Province of
Mendoza, Argentina ($35.1^\circ - 35.5^\circ$~S, $69.0^\circ -
69.6^\circ$~W, 1400~m a.s.l.). The surface array consists of
1600 water-Cherenkov detectors laid out over 3000~km$^2$ on a
triangular grid of 1.5~km spacing.  It has been in operation since 1
January 2004, increasing its size from 154 detectors up to 1600 by
June 2008. Features of the Observatory that are relevant to the present
analysis, that include data taken between 1 January 2004 and 31 December 2009, 
are outlined below.

The trigger requirement for the surface detector is based on a 3-fold
coincidence, satisfied when a triangle of neighboring stations is
triggered. A fiducial cut is applied to triggered events to ensure adequate 
containment inside the array. The cut requires that at least five active stations 
surround the station with the highest signal, and that the
reconstructed shower core be inside a triangle of active
detectors. For CR primary energies above $3\times~10^{18}$~eV, the
efficiency of this trigger chain is 100\% \cite{acceptance}. 
The exposure is determined by purely geometrical considerations, 
the uncertainty being less than 3\%. Note that analyses involving
a flux calculation, such as the measurement of the cosmic ray spectrum
\cite{paoflux,paoflux2}, use stricter fiducial cuts, which amount to a lower exposure.

The arrival directions are obtained through the differences in the
time of flight of the shower front among the triggered detectors.  The
angular resolution is defined as the angular radius around the true
cosmic ray direction that would contain 68\% of the reconstructed
shower directions. It is cross-checked using events detected
simultaneously with the fluorescence detector, {\it i.e.} hybrid events. It is better than
$0.9^\circ$ for events that trigger at least six surface stations
($E\gtrsim10$~EeV) \cite{angularresolution}.
We have tested that the angular resolution has been stable within $0.1^\circ$ during the period 
of the present analysis. 

The estimator for the primary energy is the reconstructed signal at
1000~m from the shower core, denoted $S(1000)$. The conversion from
this estimator to energy is derived experimentally through the use of
a subset of showers detected simultaneously with the fluorescence detector and
the surface array. The energy resolution is about 15\% and the absolute
energy scale has a systematic uncertainty of 22\% \cite{paoflux,paoflux2}.
We have checked the time-stability of the energy assignment by computing the fluxes in the energy
range from 10 to 55 EeV for five different periods with similar exposure. The fluxes obtained
for period I, period II, and for three equi-exposure intervals in period III
(see Table~\ref{periods} for the definition of periods I, II and III) are 0.208, 0.222, 
0.234, 0.223 and 0.226 km$^{-2}$~sr$^{-1}$~y$^{-1}$ respectively, each with an uncertainty of 
0.008 km$^{-2}$~sr$^{-1}$~y$^{-1}$, corresponding to $\sim$~1000 events in each interval.  
Given the spectral slope of 2.6 in this energy range \cite{paoflux2} and with the assumption of 
constant flux, this implies that the energy resolution of the Observatory has been stable to 
5\% over the six years of data taking.  The fluxes derived from the small number of events 
above 55 EeV are similarly constant.  

In the present analysis, we consider events recorded with the surface detector between 
1 January 2004 and 31 December 2009 with zenith angles $\theta
\le 60^\circ$ and reconstructed energy $E \ge 55$~EeV: 69 events satisfy these requirements. 
The integrated exposure for this event selection is 20,370
km$^2$~sr~y. The exposure and statistics of events in different data-taking periods are 
given in Table \ref{periods}.  
The arrival directions and energies are listed in the appendix.

\section{Update of the correlation study with AGNs in the VCV catalog\label{vcv}}

The data reported in \cite{pao1,pao2} (periods I and II in
Table~\ref{periods}) consist of 27 CR events with energy larger than
$E_{th}=55$~EeV (in the present energy calibration). These data
provided evidence for anisotropy in the arrival
directions of cosmic rays with the highest energies.

The confidence level for the rejection of the isotropic hypothesis was
established through a specific test using prescribed parameters.
Using data of period I, the values of the energy threshold, 
maximum angular separation, and maximum redshift were chosen as 
those that minimised the probability that the correlation with AGNs 
in the VCV catalog could occur by chance if the flux were isotropic.
The test was then performed using data collected subsequent
to the parameter specification by the exploratory scan. It measured the fraction of 
arrival directions that are less than $3.1^\circ$ from the position of 
an AGN within 75 Mpc in the VCV catalog. The fraction expected under the 
isotropic hypothesis is 21\%. The correlation was measured with
exactly the same reconstruction algorithms, energy calibration and
quality cuts for event selection as in the exploratory scan. 
With 6 out of 8 events correlated, the test established a 99\% confidence
level for rejecting the hypothesis that the distribution of arrival
directions is isotropic.

\begin{figure}[H]
\centering
\includegraphics[width=0.8\linewidth,angle=0]{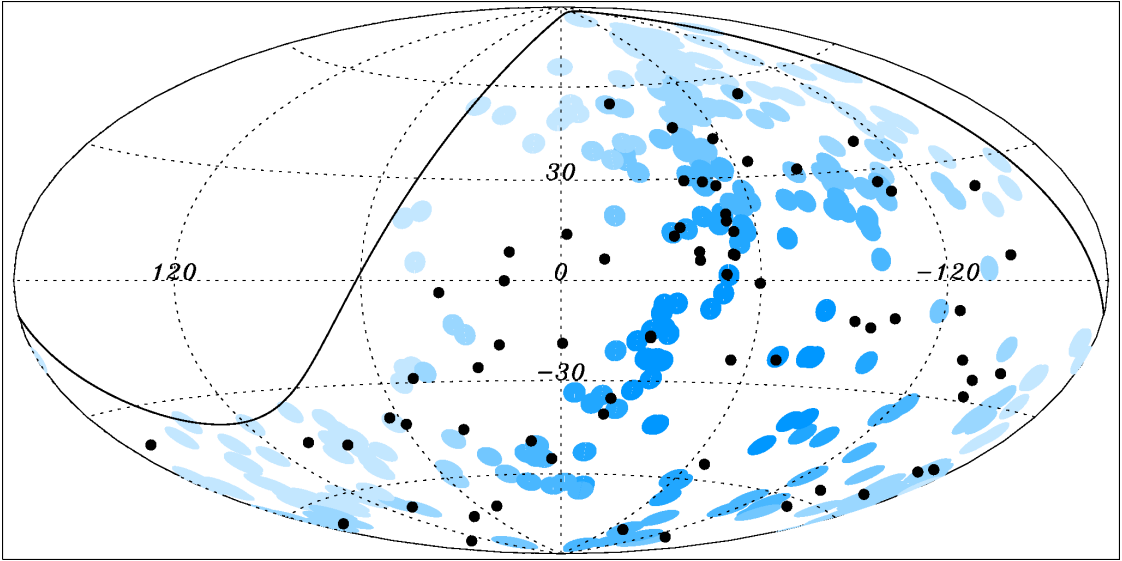}
\caption{The 69 arrival directions of
  CRs with energy $E \ge 55$~EeV detected by the Pierre Auger Observatory up to 31 December 2009
  are plotted as black dots in an Aitoff-Hammer projection of the sky in galactic coordinates. 
  The solid line represents the field of view of the Southern Observatory 
  for zenith angles smaller than $60^\circ$.
  Blue circles of radius $3.1^\circ$ are centred at the positions of the 318 AGNs in the VCV catalog that lie within 75 Mpc and that are within the field of view of the Observatory.  Darker blue indicates larger relative exposure. The exposure-weighted fraction of the sky covered by the blue circles is 21\%.}
\label{vcvmap}
\end{figure}

The number of correlations within $3.1^\circ$ between the 69 arrival
directions of CRs with $E \geq 55$~EeV detected up to 31 December 2009 and
AGNs in the VCV catalog with redshift $z \le 0.018$ are summarised in
Table~\ref{periods} and illustrated in 
Fig.~\ref{vcvmap}.\footnote{Differences with the numbers reported 
in \cite{pao1,pao2,doug} arise from small differences in the reconstruction of 
the arrival directions, as detailed in the appendix.}  
The CR events additional to those reported in
\cite{pao1,pao2} are the 42 listed for period III. Of
those 42 new arrival directions, 12 of them correlate with objects
in the VCV catalog defined by the prescribed parameters. The 
number of correlations expected by chance if the arrival directions
were isotropically distributed is 8.8.

\begin{table*}[!ht]
\caption{\label{periods}Summary of correlations within $3.1^\circ$
  between CRs with $E \geq 55$~EeV and AGNs in the VCV catalog with
  redshift $z \le 0.018$. $N$ is the number of CRs measured. $k$ is the
  number of correlating arrival directions. $k_{\rm{iso}}$ is the
  number of correlations expected by chance if the flux were
  isotropic. $P$ is the cumulative binomial probability to detect $k$
  or more correlations from an isotropic distribution. Probabilities are 
not shown for data sets which include period I
because parameters were selected to optimise the correlation in that
period.} 
        \begin{center}
                \begin{tabular}{|c|c|c|c|c|c|c|c|}
                        \hline
                        Period & Dates &Exposure&  $N$ & $k$ & $k_{\rm{iso}}$ & $P$ \\
                         & &km$^2$~sr~y &   &  &  & \\
                        \hline
                        I & 1 Jan 2004 - 26 May 2006 & 4390  & 14 &  8 & 2.9 & --  \\
                        \hline
                        II & 27 May 2006 - 31 Aug 2007 & 4500  & 13 &  9 & 2.7 & $2\times 10^{-4}$ \\
                        \hline
                        III & 1 Sept 2007 - 31 Dec 2009 & 11480  & 42 &  12 & 8.8 & 0.15 \\
                        \hline
                        Total & 1 Jan 2004 - 31 Dec 2009 & 20370  & 69 &  29 & 14.5 & --  \\
                        \hline
                        II+III & 27 May 2006 - 31 Dec 2009& 15980 &  55 &  21 &  11.6 & $\mathbf{3\times 10^{-3}}$ \\
                        \hline
                \end{tabular}
        \end{center}
\end{table*}

\begin{figure}[H]
\centering
\includegraphics[width=0.75\linewidth,angle=0]{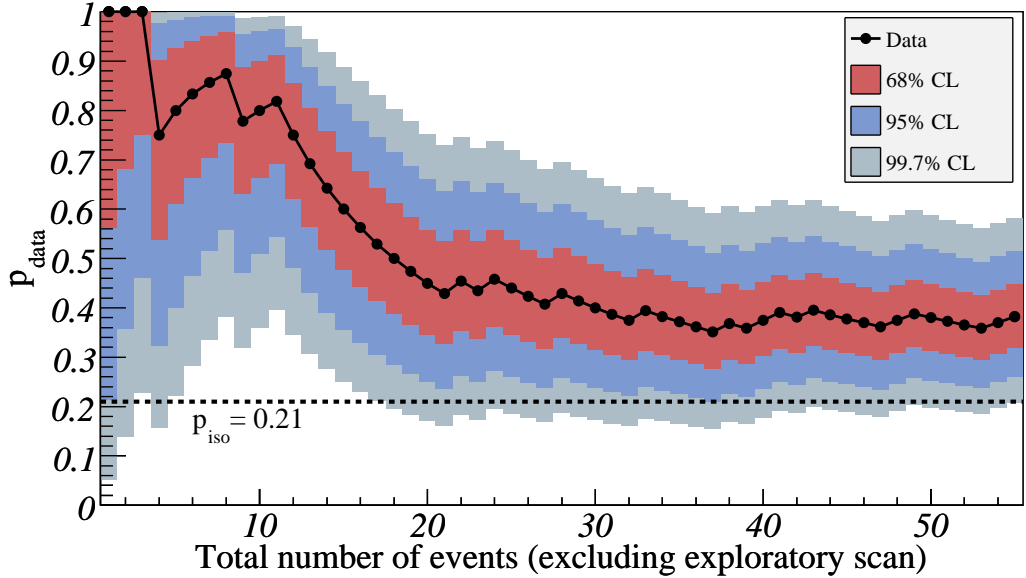}
\caption{The most likely value of the degree of correlation
  $p_{\rm data}=k/N$ is plotted with black dots as a function of the
  total number of time-ordered events (excluding those in period I). 
  The 68\%, 95\% and 99.7\% confidence level intervals
  around the most likely value are shaded.  The horizontal
  dashed line shows the isotropic value $p_{\rm iso} = 0.21$.  The
  current estimate of the signal is $(0.38^{+0.07}_{-0.06})$.}  
\label{seq}
\end{figure}

The updated estimate of the degree of correlation must include periods
II and III only, because the parameters were chosen to maximise
the correlation in period I. In Fig.~\ref{seq}
we plot the degree of correlation ($p_{\rm data}$) with objects in the
VCV catalog as a function of the total number of time-ordered events
observed during periods II and III. 
For each additional event the most
likely value of $p_{\rm data}$ is $k/N$ (number correlating divided by
the cumulative number of arrival directions). 

The confidence level intervals
in the plot contain 68.3\%, 95.45\% and 99.7\%
of the posterior probability for $p_{\rm data}$ given the measured values of $k$ and $N$. 
The posterior probability distribution is 
\nobreak{$p_{\rm data}^k(1-p_{\rm data})^{N-k}(N+1)!/k!(N-k)!$,} 
corresponding to a binomial likelihood with a flat prior.
The upper and lower limits in the confidence intervals are chosen such that the
posterior probability of every point inside the interval is higher than 
that of any point outside.
The amount of correlation observed has decreased from
$(69^{+11}_{-13})\%$, with 9 out of 13 correlations measured in
period II, to its current estimate of $(38^{+7}_{-6})\%$, based on 21
correlations out of a total of 55 events in periods II and III.

The cumulative binomial probability that an isotropic flux would yield
21 or more correlations is $P=0.003$. This updated measurement with 55
events after the initial scan is {\it a posteriori}, with no
prescribed rule for rejecting the hypothesis of isotropy as in
\cite{pao1,pao2}.  No unambiguous confidence level for anisotropy 
can be derived from the probability $P=0.003$. 
$P$ is the
probability of finding such a correlation assuming isotropy.
It is not the probability of isotropy given such a correlation.

We note that 9 of the 55 events detected in periods II and III are
within $10^\circ$ of the galactic plane, and none of them correlates
within $3.1^\circ$ with the astronomical objects under
consideration. Incompleteness of the VCV catalog due to obscuration by
the Milky Way or larger magnetic bending of CR trajectories along the
galactic disk are potential causes for smaller correlation of arrival
directions at small galactic latitudes. If the region within
$10^\circ$ of the galactic plane is excluded the correlation is $(46\pm 6)\%$ 
(21 correlations out of 46 events), while 24\% is the chance expectation 
for an isotropic flux.\footnote{The choice of the 
size of the region excluded has some arbitrariness. We used $12^\circ$ 
in \cite{pao1,pao2}. We use $10^\circ$ here for uniformity with the analysis 
of the 2MRS catalog in section \ref{othercatalogs}.} 

It has not escaped our notice that the directions of the 5 most
energetic events are not part of the fraction of events that correlate
with objects in the VCV catalog.

Additional monitoring of the correlation signal with this set of
astronomical objects can also be found in \cite{doug}. 
Further studies of the correlation exploring other parameters are currently in progress.
One conjecture often made in the literature (see e.g.\ \cite{ginzburg,biermann} and references therein) is that powerful radiogalaxies are the most promising contenders for UHECR acceleration, along with gamma-ray bursts. The analysis of directional correlations of UHECRs with positions of AGNs from the VCV catalog discussed here does not account for any differences among those AGNs. Thus, a logical next step with respect to \cite{pao1,pao2} would consider the AGN radio luminosity given in the VCV catalog as a fourth scan parameter to find a threshold in radio luminosity above which the directional correlation starts to increase. Such a scan has been performed with a subset of the data and the signal evolution with those parameters is being monitored since, similarly as presented here for all AGNs of the VCV. These results will be reported elsewhere.

The HiRes collaboration has reported \cite{hiresagn} an absence of a correlation with
AGNs of the VCV catalog using the parameters of the Auger prescribed
test.  They found two events correlating out of a set of 13
arrival directions that have been measured stereoscopically
above an energy which they estimated to be the same as the Auger 
prescribed energy threshold.  The $38\%$ correlation measured by Auger 
suggests that approximately five arrival directions out of 13 HiRes directions
should correlate with an AGN position.  The difference between 2 and 5
does not rule out a $38\%$ correlation in the northern hemisphere that
is observed by the HiRes detector.  Also, it is not necessarily
expected that the correlating fraction should be the same in both
hemispheres.  The three-dimensional AGN distribution is not uniform,
and the VCV catalog itself has different level of completeness in the two
hemispheres.  In addition, comparison of results between the two
observatories is especially challenging in this situation because the
energy cut occurs where the GZK suppression has steepened the already
steep cosmic ray spectrum.  A small difference in the threshold energy
or a difference in energy resolution can strongly affect the
measurement of a correlation that exists only above the threshold.

It is worth mentioning that while the degree of
correlation with the parameters of the test updated here has decreased
with the accumulation of new data, a re-scan of the complete data set
similar to that performed in Ref.~\cite{pao2} does not lead to a much more
significant correlation for other values of the parameters.  The
largest departure from isotropic expectations in the scan actually
occurs for the same energy threshold $E_{th}=55$~EeV and maximum
redshift $z\le 0.018$. There is a spread in the angular scales over
which the correlation departs from isotropic expectations. This issue
will be examined in section~\ref{othercatalogs}, where we explore the correlation with
other sets of nearby extragalactic objects, described by catalogs more
uniform than the VCV compilation.

There is now available a more recent
version of the VCV catalog~\cite{vcv13}.
Conclusions are similar if the arrival directions are compared to
the distribution of objects in this latest version.

\section{Examination of the arrival
directions in relation to other catalogs\label{othercatalogs}}

As noted in \cite{pao1}, ``the correlation that we observe with nearby
AGNs from the VCV catalog cannot be used alone as a proof that AGNs are
the sources. Other sources, as long as their distribution within the
GZK horizon is sufficiently similar to that of the AGNs, could lead to
a significant correlation between the arrival directions of cosmic
rays and the AGNs positions.'' It is therefore appropriate to
investigate the arrival directions of this data set with respect to
other scenarios for cosmic ray sources in the local universe.

It is important to note that all of these studies are made {\it a posteriori}.  
None of the results can be used to derive unambiguously a confidence level for anisotropy.
The single-trial VCV test that was prescribed in 2006 resulted in 99\%
confidence that the flux of cosmic rays is not isotropic \cite{pao1,pao2}.
The $P$-value 0.003 reported in section \ref{vcv} does not increase
confidence in anisotropy beyond what was reported in \cite{pao1,pao2}.  
With the currently estimated correlation fraction of
38\%, a $5\sigma$ significance ($P<6\times 10^{-7}$) will require 165 events
subsequent to period I, and that larger data set will not be available
for at least another four years.  In the meantime, it is natural to explore the
present data set to see if scenarios other than the simple VCV
correlation are supported by the current set of arrival directions.
Even when (or if) a $5\sigma$ deviation from isotropy is established via
the VCV correlation, it will be important to determine the best
astrophysical interpretation for it.  At that time, it could be
interesting to test if any of the scenarios investigated here may have
acquired additional supporting evidence.

The same minimum energy of CRs will be used for these exploratory studies as
was prescribed in 2006 for the VCV test.  The idea is to examine the
same set of 69 arrival directions using alternative models.  Each
model has its own set of relevant parameters, and those will be
separately tuned.  In the prescribed VCV test there were three important
parameters.  One was the minimum energy that defines the set of
arrival directions.  The other two were the correlation angle
($\psi=3.1^\circ$) and the maximum AGN redshift ($z_{max}=0.018$) which
pertain to the model.  It would be possible to optimise the minimum
energy cut also for every scenario, as was done prior to prescribing
the VCV test.  For the studies here, however, the data set will be
kept the same.  It includes all recorded events above 55 EeV. 
By including
period I, which was used to optimise the energy cut for the VCV
correlation in that period, scenarios similar to the prescribed VCV
model could be favored.
The effect of excluding the events used in the exploratory scan,
that are strongly correlated with VCV objects, will
be analysed.

In what follows we examine the present data set of arrival directions with regard to their
correlation with different populations of nearby extragalactic objects: 
galaxies in the 2MRS catalog and AGNs detected by Swift-BAT.
We choose these sets of objects as examples of astrophysical scenarios 
worthy of examination. We have reported additional explorations (such as the correlation with 
galaxies in the HI Parkes All Sky Survey \cite{HIPASS,HIPASS2}) in \cite{julien}. 

The 2MRS catalog is the most densely sampled all-sky redshift survey to date. It is a compilation
provided by Huchra et al.~\cite{2MRS} of the redshifts of the $K_\mathrm{mag}<11.25$ 
brightest galaxies from the 2MASS catalog~\cite{2MASS}. It contains approximately 13000 galaxies 
within $100$~Mpc, and 22000 within $200$~Mpc.  It provides an unbiased 
measure of the distribution of galaxies in the local universe, out to a mean redshift
of $z = 0.02$, and to within $10^\circ$ of the Galactic plane. To avoid biases due to its incompleteness 
in the galactic plane region, we exclude from all analyses involving this catalog 
galaxies (as well as CR arrival directions) with galactic latitudes $|b|<10^\circ$. 

The Swift-BAT hard X-ray catalog \cite{SWIFT22} is the product of the most sensitive all-sky survey in the hard X-ray band.  We use the 58-month version of the Swift-BAT survey \cite{SWIFT58}. A sample of AGNs selected from the hard X-ray band reduces the bias due to absorption that affects an optical selection. We consider for the present analysis all Seyfert galaxies, beamed AGNs, and galaxies likely to be AGN but with no confirmed nuclear activity in the optical spectrum. There are 189 of them within approximately $100$~Mpc, and 373 within approximately $200$~Mpc. 

\subsection{Cross-correlation of cosmic rays and nearby extragalactic
objects\label{xcorr}}

We report the result of a direct cross-correlation
analysis between arrival directions of CRs and positions of the objects in
the 2MRS and Swift-BAT catalogs that lie within 200~Mpc. 
Each CR arrival direction forms a pair with every object in the catalogs.
For the cross-correlation estimator, we use the fractional excess
(relative to the isotropic expectation) of pairs having angular
separations smaller than any angle $\psi$.  This is given by $n_{\rm
  p}(\psi)/n_{\rm p}^{\rm iso}(\psi)-1$, where $n_p(\psi)$ denotes the
number of pairs with separation angle less than $\psi$. Departures
from isotropy are higher if arrival directions correlate with regions
with larger density of objects. 

\begin{figure}[H]
\centerline{\hfill\includegraphics[angle=-90,width=0.5\linewidth]{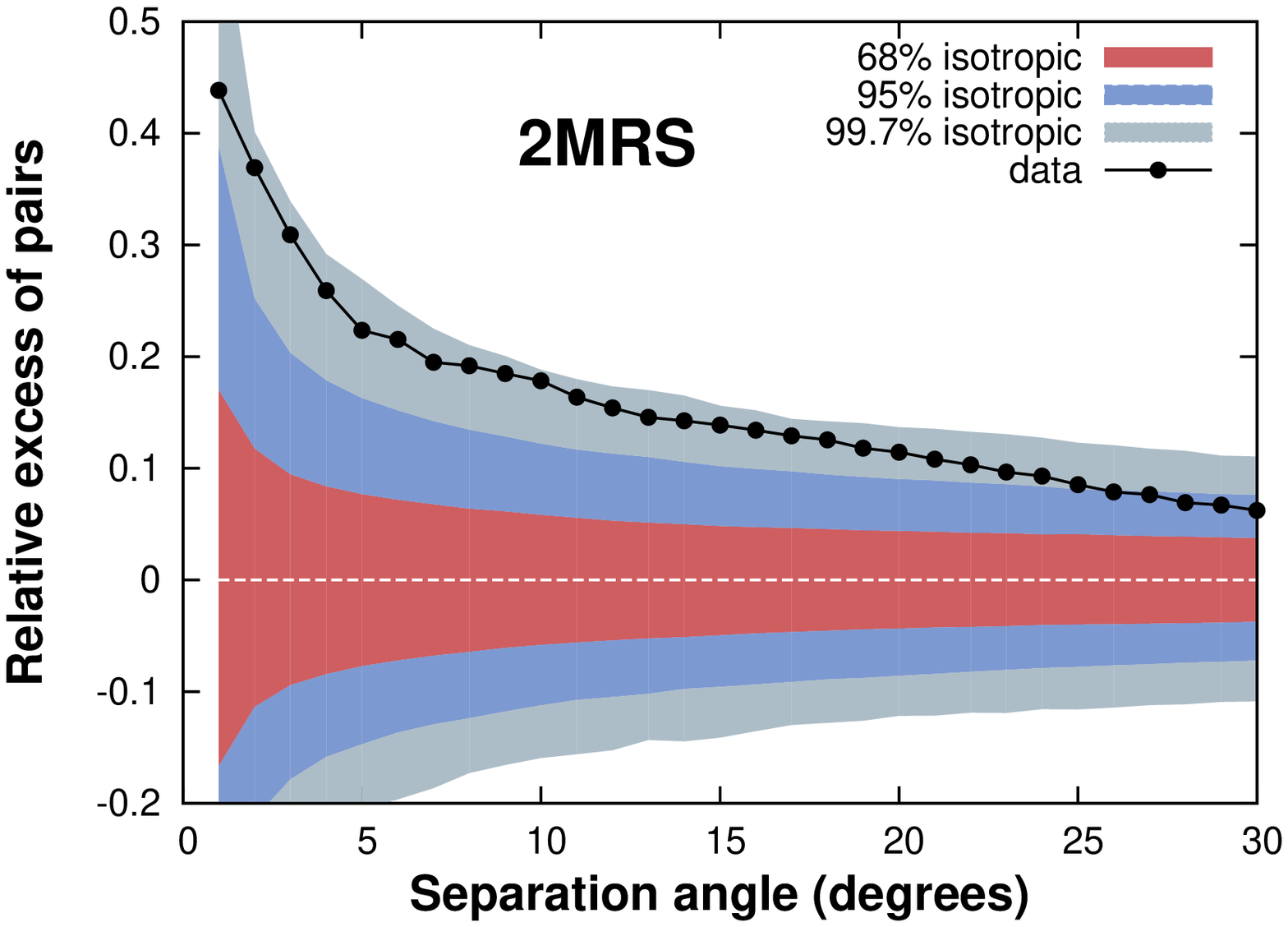}
\hfill\includegraphics[angle=-90,width=0.5\linewidth]{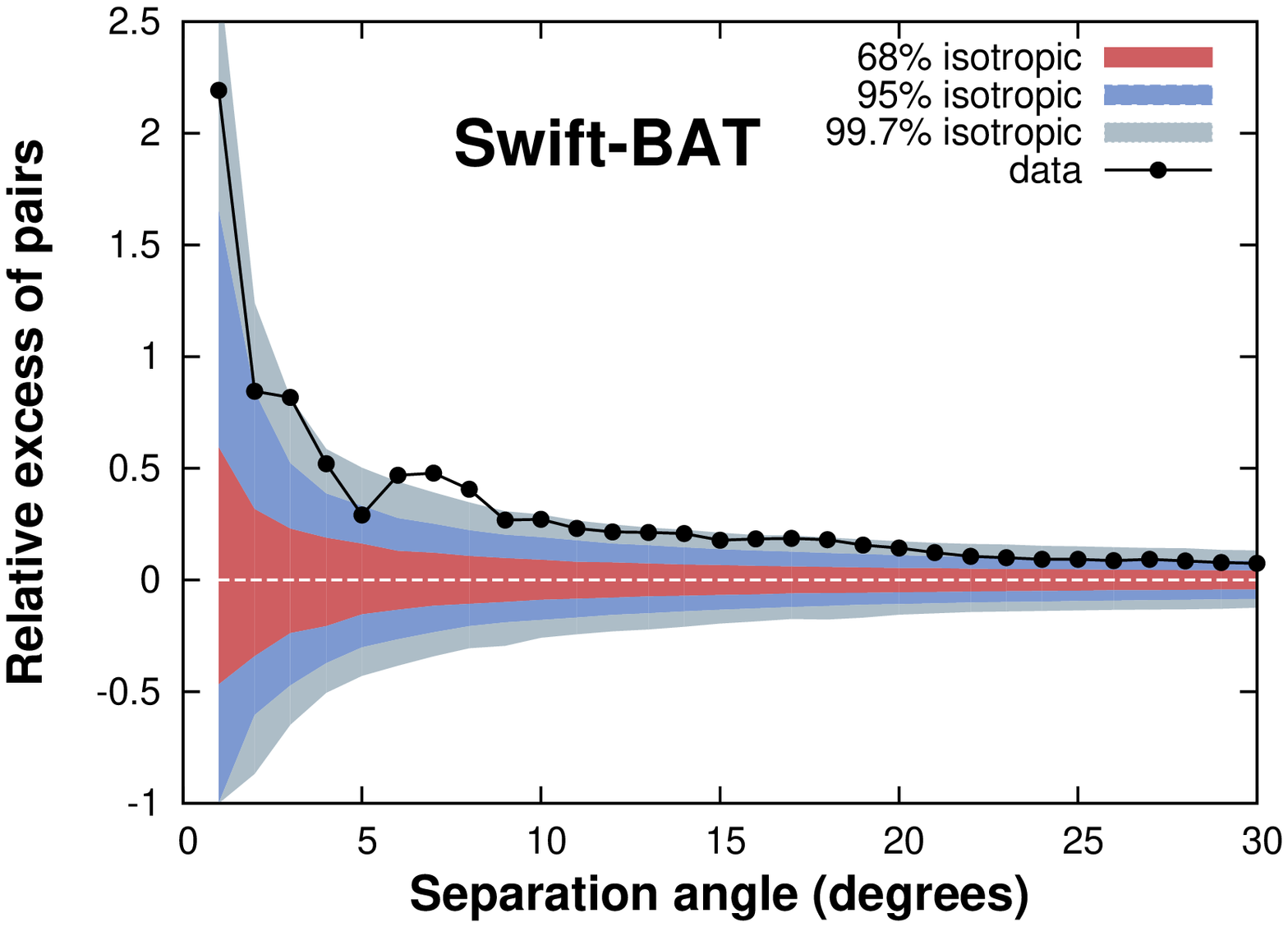}\hfill}
\centerline{\hfill\includegraphics[angle=-90,width=0.5\linewidth]{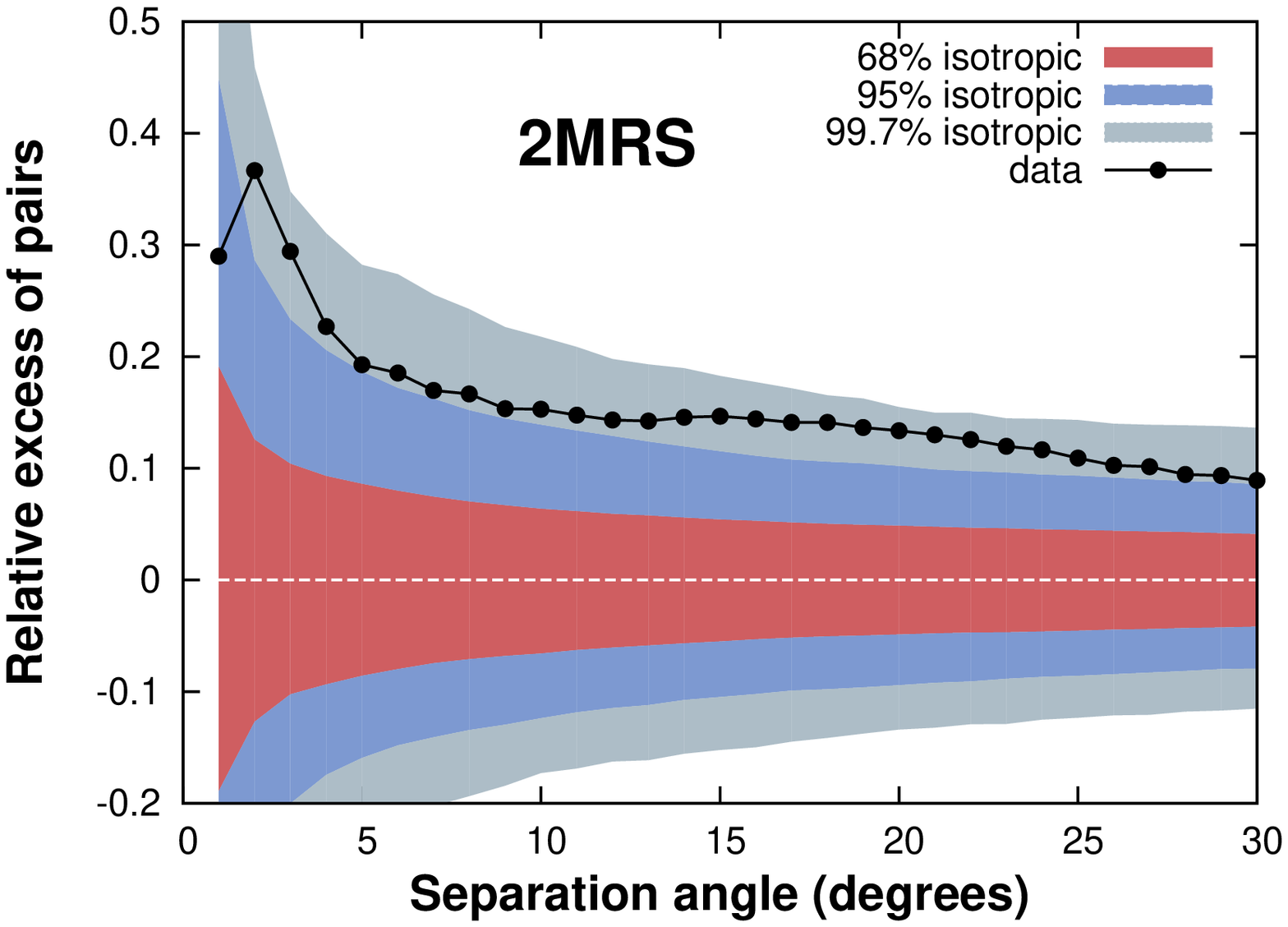}
\hfill\includegraphics[angle=-90,width=0.5\linewidth]{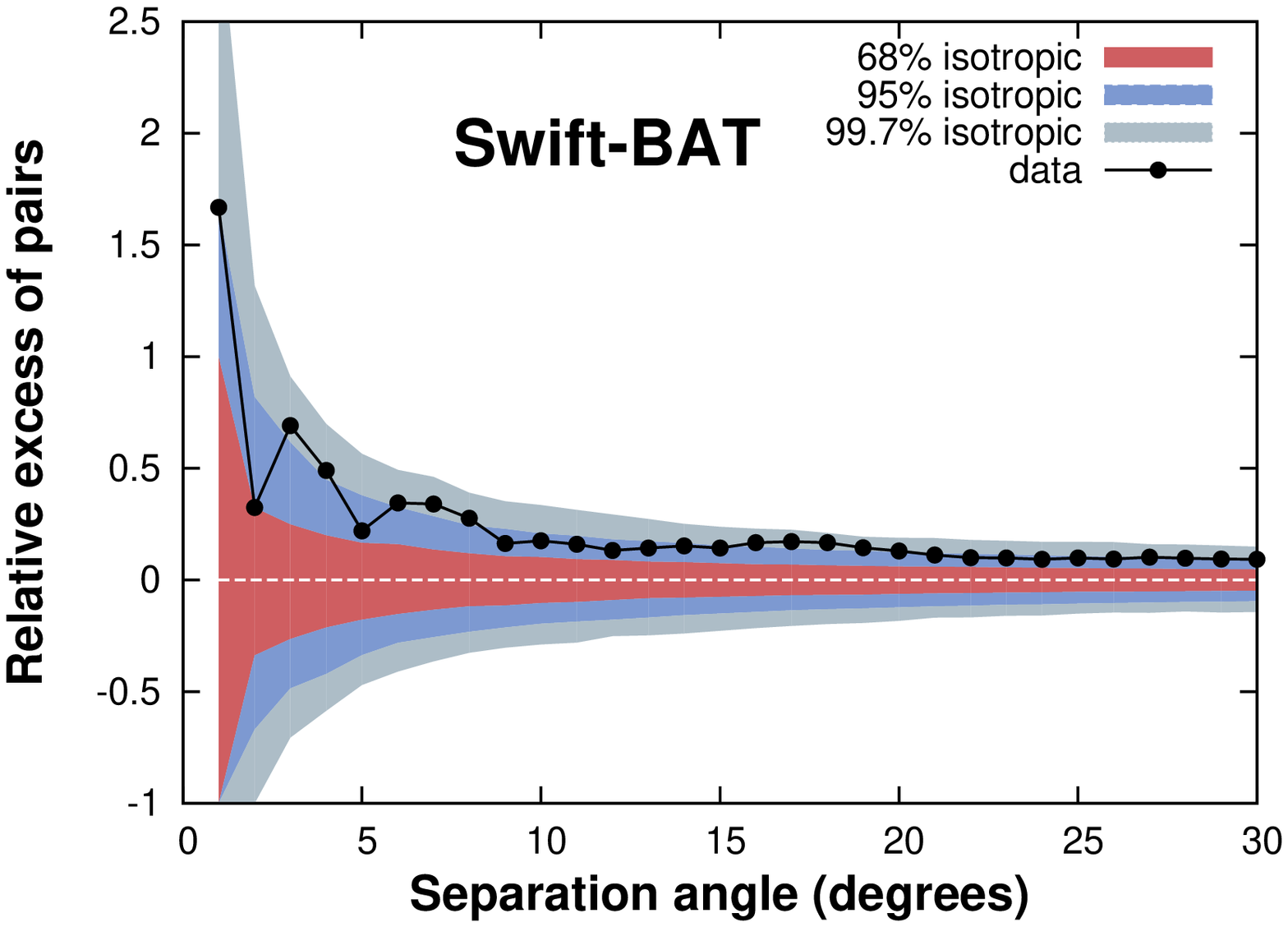}\hfill} 
\caption{Cross-correlation between the arrival directions of CRs
  measured by the Pierre Auger Observatory with $E \ge 55$~EeV and
  positions of 2MRS galaxies (left)
and Swift-BAT AGN (right) that lie within $200$~Mpc. In the case of 2MRS
  galactic latitudes (both of galaxies and CRs) are restricted to $|b|>10^\circ$. 
The plots in the top panels use all CRs with $E \ge 55$~ EeV. The plots in the bottom panels exclude data
collected during period I in Table \ref{periods}, that were used to choose the energy threshold
and redshift that maximized the correlation with VCV objects in that period. The bands correspond to the 68\%, 95\% and 99.7\% 
dispersion expected for an isotropic flux.} \label{fig:XC3}
\end{figure}

We plot in Fig.~\ref{fig:XC3} the relative excess of pairs using data (black dots)
in the case of 2MRS galaxies (left)
and Swift-BAT AGNs (right). 
The bands in the plot contain the dispersion in 68\%, 95\%, and 99.7\% 
of simulated sets of the same number of events
assuming isotropic cosmic rays. The top panels plot the results using all the arrival directions 
of CRs with $E \ge 55$~EeV collected between 1 January 2004 and 31 December 2009: 69 CR events 
in the case of correlation with 
Swift-BAT AGNs, and 57 CR events in the case of 
correlation with galaxies in the 2MRS catalog (for which galactic latitudes $|b|<10^\circ$ 
were excluded). 
The bottom panels plot the results excluding the arrival directions 
of CRs collected during period~I in Table~\ref{periods}, which were used to optimise 
the energy cut for the VCV correlation in that period: 55 CRs are used in the case 
of correlation with 
Swift-BAT AGNs, and 46 CRs in the case of correlation with galaxies 
in the 2MRS catalog. Features in the plots are comparable if period I is excluded.

We observe correlation in excess of isotropic expectations in all cases. Note however that the existence
of cross-correlation does not imply
that the arrival directions are distributed in the sky in the same
manner as the objects under consideration.

The catalogs of astronomical objects that were used here are flux-limited sets. A similar analysis confronting the
arrival directions with a volume-limited subsample of the 2MRS catalog was reported in Ref.~\cite{julien}.

\subsection{Statistical tests on smoothed density maps\label{maps}}

\subsubsection{Smoothed density maps}

We test some specific models for the origin of the highest energy CRs
based on the astronomical objects in the catalogs considered in
the previous section. We build the probability maps of arrival
directions of CRs expected from these objects weighted by their flux 
at the electromagnetic wavelength relevant in the respective survey 
and by the attenuation factor expected from the GZK effect.  
Maps are constructed by the
weighted superposition of Gaussian distributions centred at each
object position with a fixed angular width $\sigma$. For each
model, the density map has two free parameters: the smoothing angle
$\sigma$ and an isotropic fraction $f_{\rm iso}$.  The smoothing angle
serves to account for typical (but unknown) magnetic deflections in
the CR trajectories. The addition of an isotropic fraction is a way to
account for CR trajectories that have been bent by wide angles due to
large charges and/or encounters with strong fields.  

A large
isotropic fraction could also indicate that the model is not using a
set of objects that includes all of the contributing CR sources.
The missing flux contributed by the relatively fainter sources below the
flux-limit of a survey can be estimated if a model for the luminosity
distribution is assumed. For instance, in a flux-weighted model based
on objects with a luminosity distribution described by a Schechter
function \cite{schechter} in a survey with characteristic depth of
$130$~Mpc, account taken of the GZK effect with an energy threshold of
$60$~EeV, the fraction of missing flux is estimated to be of the order
of 35\%~\cite{bariloche}.  The faint sources are not expected to be
isotropically distributed, and thus an isotropic fraction may not be
an accurate representation for the distribution of that missing flux. An
alternative to the addition of an isotropic fraction, when selection
effects as a function of distance are known, is to divide the observed
density of galaxies at a given distance by the selection function
\cite{wax,cuoco}. A possible drawback of this approach is that one
assigns the unobserved galaxies to the same locations where bright
galaxies are observed, and this may introduce a bias.  

We will not assume specific values for the isotropic fraction and
smoothing angles introduced into the models, but rather use the data
to determine the best fit values of these parameters.

The smoothed maps are described by a function $F({\bf \hat{n}})$,
such that its value in a given direction $\bf
\hat{n}$ is proportional to the probability of detecting a
cosmic ray in that direction, according to the model.  We write the
function $F({\bf \hat{n}})$ as: \begin{equation} F({\bf \hat{n}})=
\frac{\varepsilon({\bf \hat{n}}) \mu({\bf \hat{n}})}{I} \, \left[
\frac{f_\mathrm{iso}}{ \Omega} + (1-f_\mathrm{iso})\frac{ \phi({\bf
\hat{n}})} {\langle\phi \rangle}\right]. \label{Fc} \end{equation} The
two terms in the sum between brackets are the isotropic component
(parameterised by $f_\mathrm{iso}$) and the contribution from the
astronomical objects.  $\Omega=\int\mathrm{d}\Omega\mu({\bf \hat{n}})$
is the solid angle subtended by the region of the sky covered by the
survey. $\mu({\bf \hat{n}})$ is the mask function of the catalog, that
vanishes in the regions of the sky that must be removed (such as that
along the galactic plane in the case of the 2MRS catalog) and
is unity elsewhere. The flux coming from the objects in the catalog is
represented by the term \begin{equation} \phi({\bf \hat{n}})
=\sum_{i=1}^{N_\mathrm{cat}} w(z_i) \; e^{-\frac{d({\bf
\hat{n}_i},{\bf \hat{n}})^2}{2\sigma^2}} \label{phi} \end{equation}
where $d({\bf \hat{n}_i},{\bf \hat{n}})$ is the angle between the
direction of the source ${\bf \hat{n}_i}$ and the direction of
interest ${\bf \hat{n}}$.  The sum extends over all objects in the
catalog, $N_\mathrm{cat}$. The free parameter $\sigma$ enables us to
take the angular resolution of the Observatory into account and 
the deflections experienced by cosmic rays under the
simplifying method of a gaussian smoothing.  A weight $w(z_i)$ is
attributed to the $i$th source located at redshift $z_i$. We assume a
weight proportional to the flux $\phi_i$ of the source, measured in a
given range of wavelengths (X-rays for Swift-BAT
and near IR for 2MRS). We multiply it by an attenuation factor due to
the GZK suppression, evaluated as the fraction of the events produced
above a given energy threshold which are able to reach us from a
source at a redshift $z$ with an energy still above that same
threshold \cite{bariloche}. We use the GZK suppression factor that corresponds 
to a proton composition. The suppression is comparable for
iron nuclei but is stronger for intermediate mass nuclei. The flux in
Eq.~\ref{Fc} is divided by its average
$\langle\phi\rangle=\int\mathrm{d}\Omega\mu({\bf \hat{n}})\phi({\bf
\hat{n}})$ for normalization. The term in front of the brackets in
Eq.~\ref{Fc} is an overall normalization. $\varepsilon({\bf \hat{n}})$
is the relative exposure of the Pierre Auger Observatory, derived
analytically from geometric considerations. The constant $I$
is chosen such that the integral of $F({\bf \hat{n}})$ is equal to unity.

\begin{figure}[H]
\centerline{\includegraphics[width=0.45\linewidth,angle=0]{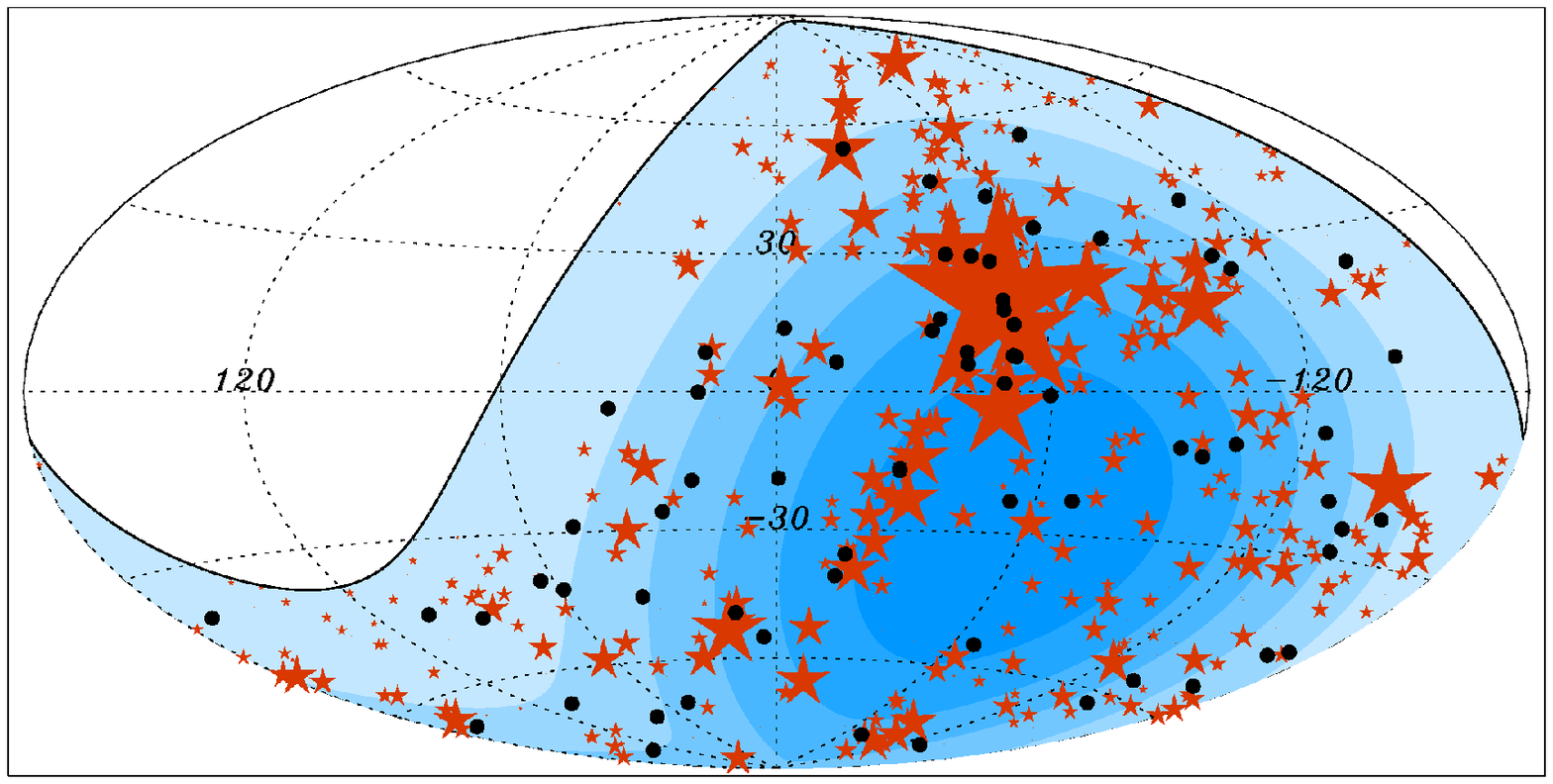}\hfill
\includegraphics[width=0.5\linewidth,angle=0]{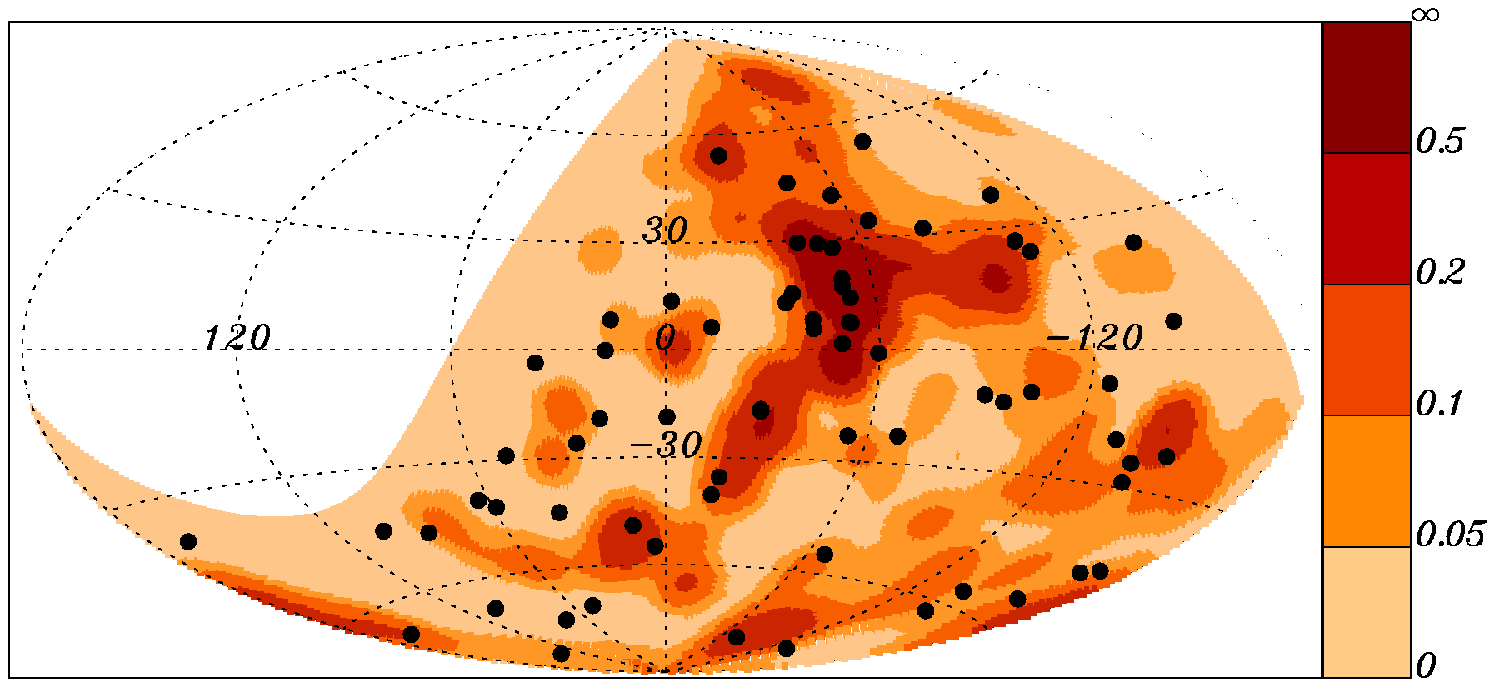}}
\caption{Left: Sky map in galactic coordinates with the AGNs of the
  58-month Swift-BAT catalog plotted as red stars with area
  proportional to the assigned weight. 
  The solid line represents the field of view of the Southern Observatory.
  Coloured bands have equal integrated exposure, and darker background
  colours indicate larger relative exposure.  
  Right: density map derived from the
  map to the left, smoothed with an angular scale
  $\sigma=5^\circ$. The 69 arrival directions of
  CRs with energy $E \ge 55$~EeV detected with the Pierre Auger Observatory 
  are plotted as black dots.} \label{swiftweighted}
\end{figure}

We illustrate in Fig.~\ref{swiftweighted} the construction of the
smoothed maps with the Swift-BAT catalog of AGNs.  The red stars on the 
left panel of Fig.~\ref{swiftweighted} are centred at the positions of
the AGNs, and the area of each star is
proportional to the weight of its AGN, determined by the X-ray flux, the
relative exposure of the Observatory, and the GZK effect.  

\begin{figure}[H]
\centerline{
\includegraphics[width=0.5\linewidth,angle=0]{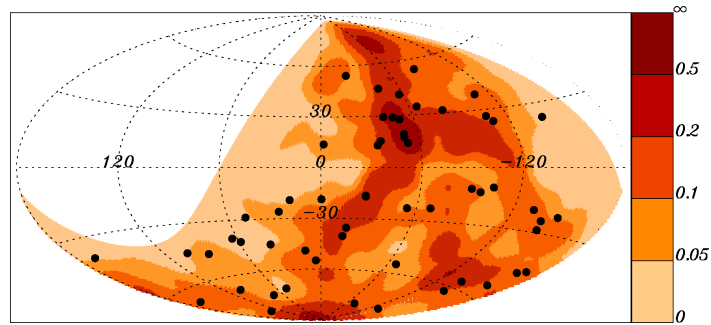}} 
\caption{\label{probMaps}Cosmic ray density map for the flux-weighted
2MRS galaxies,
smoothed with an angular scale $\sigma=5^\circ$. The black dots are the 
arrival directions of the CRs with energy $E \ge 55$~EeV detected with the Pierre Auger Observatory. 
Galactic latitudes are restricted to $|b|>10^\circ$, 
both for galaxies and CR events.}
\end{figure}

The corresponding
density map is shown on the right panel of the same figure,
smo\-oth\-ed with an angular scale $\sigma=5^\circ$. No isotropic
fraction is built into this map to better illustrate the features of
the objects in the catalog.  We show the density map obtained for the 2MRS catalogue 
in Fig.~\ref{probMaps}. Common features can be seen in the two maps. 

\subsubsection{Likelihood test}

For each model and for different values of the smoothing angle $\sigma$ and 
isotropic fraction $f_{\rm iso}$  we evaluate the log-likelihood of the data sample: 

\begin{equation}
\mathcal{LL}=\sum_{k=1}^{N_\mathrm{data}} \mathrm{ln}F({\bf
\hat{n}_k}), 
\end{equation} 
where ${\bf \hat{n}_k}$ is the direction of the $k$th event.

\begin{figure}[H]
\centerline{\hfill\includegraphics[width=0.35\linewidth]{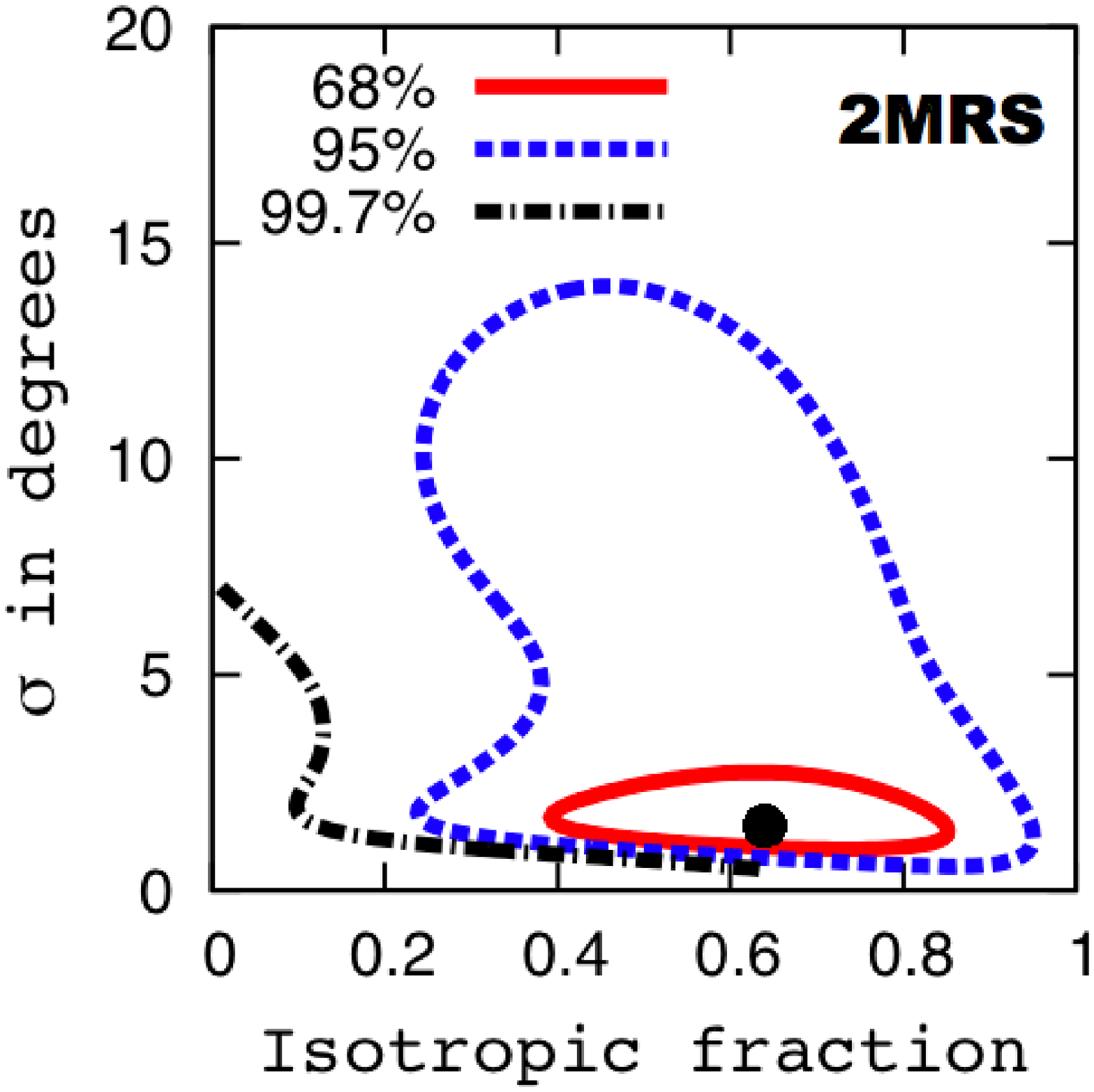}
\hfill\includegraphics[width=0.35\linewidth]{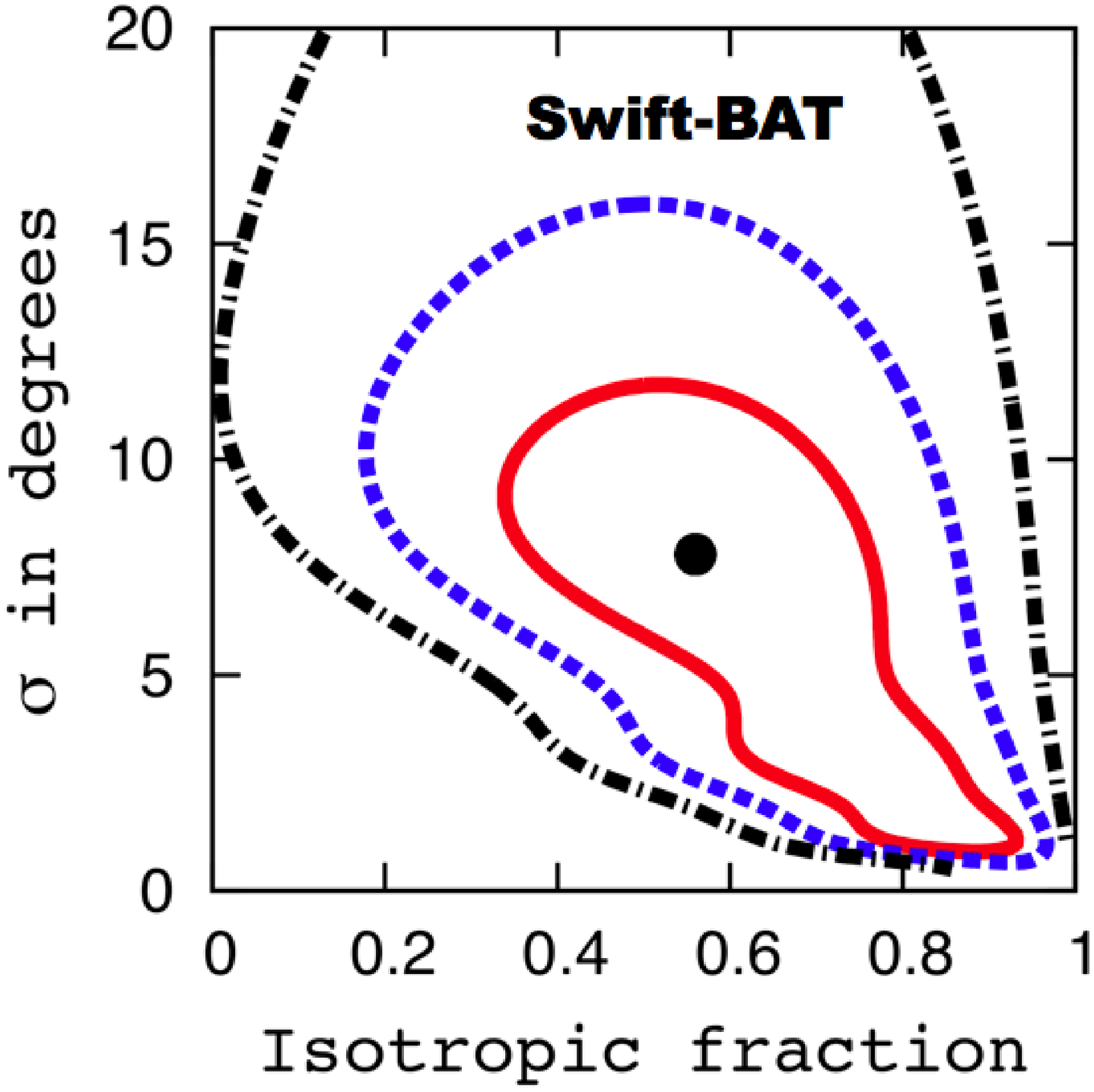}\hfill}
\centerline{\hfill\includegraphics[width=0.35\linewidth]{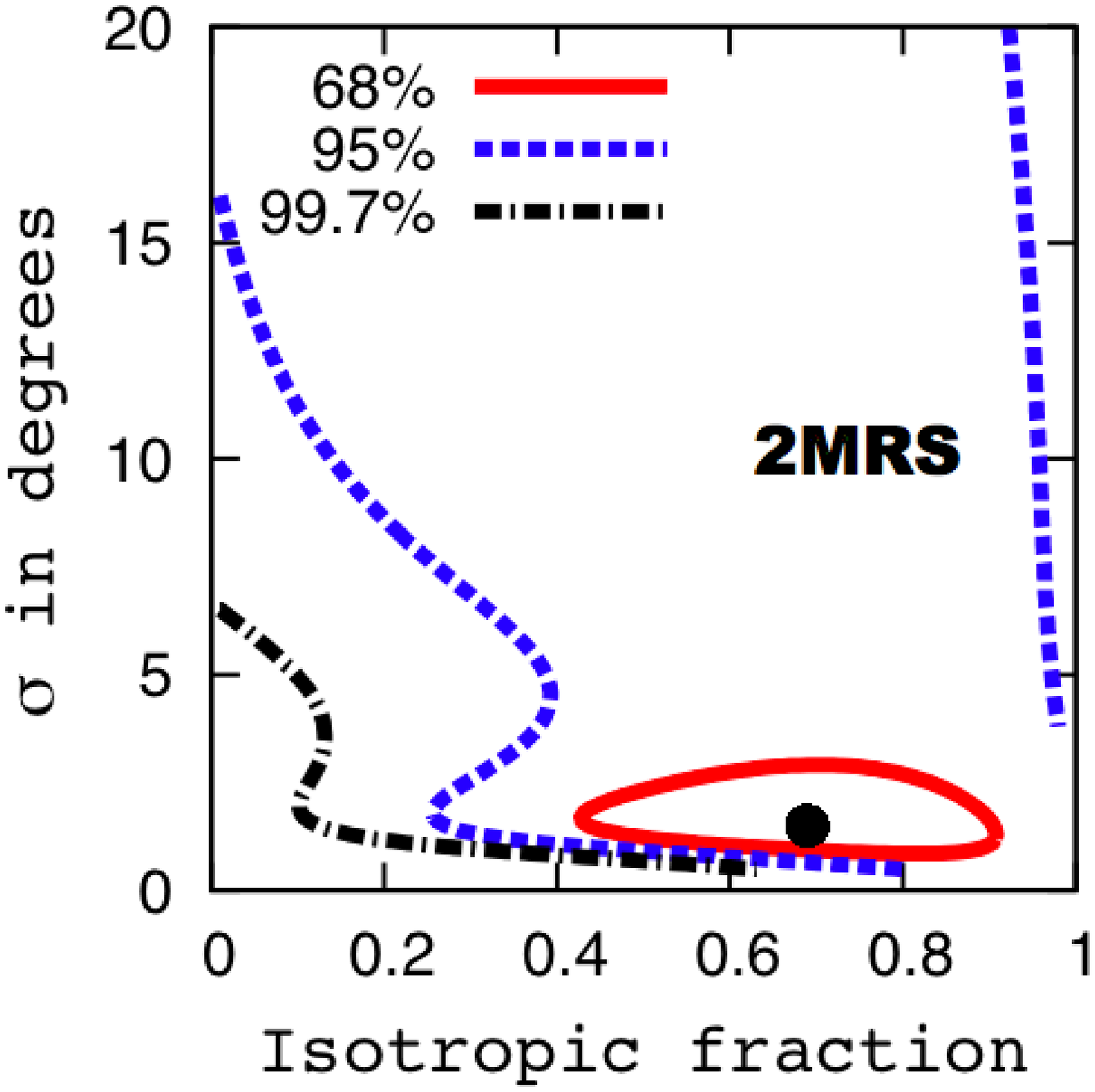}
\hfill\includegraphics[width=0.35\linewidth]{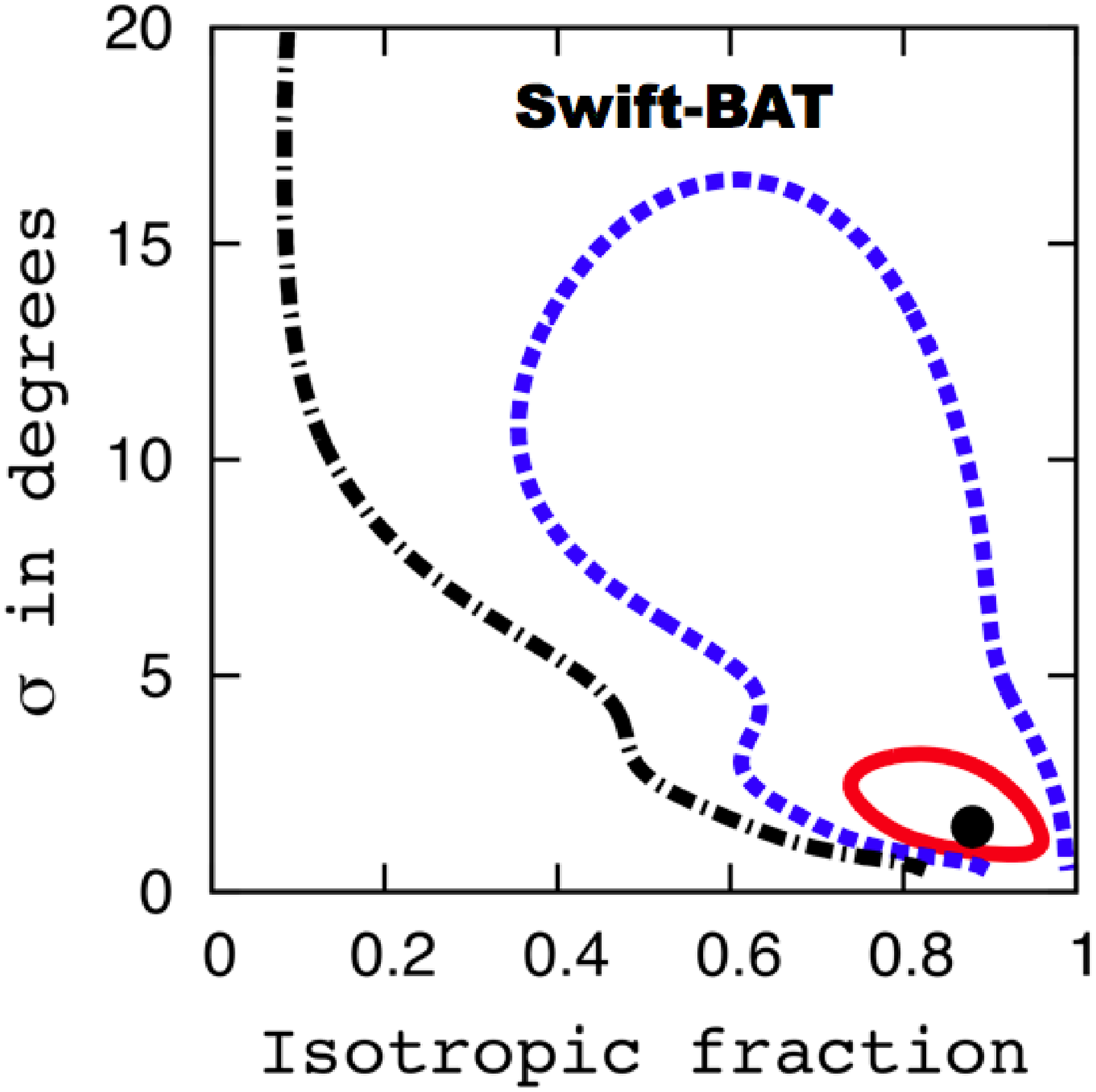}\hfill}
\caption{\label{fig:LL_cont} Confidence intervals for the parameters 
$(\sigma,f_{\rm iso})$ derived from the
likelihood function using the arrival directions of CRs with 
$E \ge 55$~EeV for the 
two models considered: 2MRS
  galaxies (left) and Swift-BAT AGNs (right).  The 
 pair of parameters that maximise the likelihood is indicated by a black dot.
The plots in the top panels use all data. The plots in the bottom panels exclude data
collected during period I in Table \ref{periods}, that were used to choose the energy threshold
that maximized the correlation with VCV objects in that period.
 In the case of 2MRS
  galactic latitudes (both of galaxies and CRs) are restricted to $|b|>10^\circ$. }
\end{figure}

We consider the models based on 2MRS 
and Swift-BAT objects weighted by their flux in the respective wavelength. 
The top panels in Fig.~\ref{fig:LL_cont} plot the results using all the arrival directions
of CRs with $E \ge 55$~EeV. 
The bottom panels
plot the results excluding the CRs collected during period I in Table~\ref{periods}, 
which were used to optimise 
the energy cut for the VCV correlation in that period. 
The best-fit values
of ($\sigma,f_{\rm iso}$) are those that
maximise the likelihood of the data sample, and are indicated
by a black dot. Contours of 68\%, 95\%, and 99.7\% confidence intervals are 
shown. The best-fit values of ($\sigma,f_{\rm iso}$) are $(1.5^\circ, 0.64)$ for 2MRS
and $(7.8^\circ,0.56)$ for Swift-BAT using all data. With data in
period I excluded the best-fit parameters are 
$(1.5^\circ, 0.69)$ for 2MRS
and $(1.5^\circ,0.88)$ for Swift-BAT.
These values are not strongly constrained with the present statistics. 
Notice for instance that the best-fit value of $f_{\rm iso}$ for the Swift-BAT model increases 
from 0.56 to 0.88 and $\sigma$ decreases from $7.8^\circ$ to $1.5^\circ$ if data in period I is excluded. 
More data is needed to discern if it is the correlation on small angles of a few events with
the very high-density regions of this model (such as the region in the direction to the radiogalaxy Centaurus A, 
the object with the largest weight in Fig.~\ref{swiftweighted}) that masks a potentially larger correlating fraction 
(hence a smaller $f_{\rm iso}$) over larger angular scales.

\begin{figure}[H]
   \centerline{\hfill\includegraphics[width=0.35\linewidth,angle=-90]{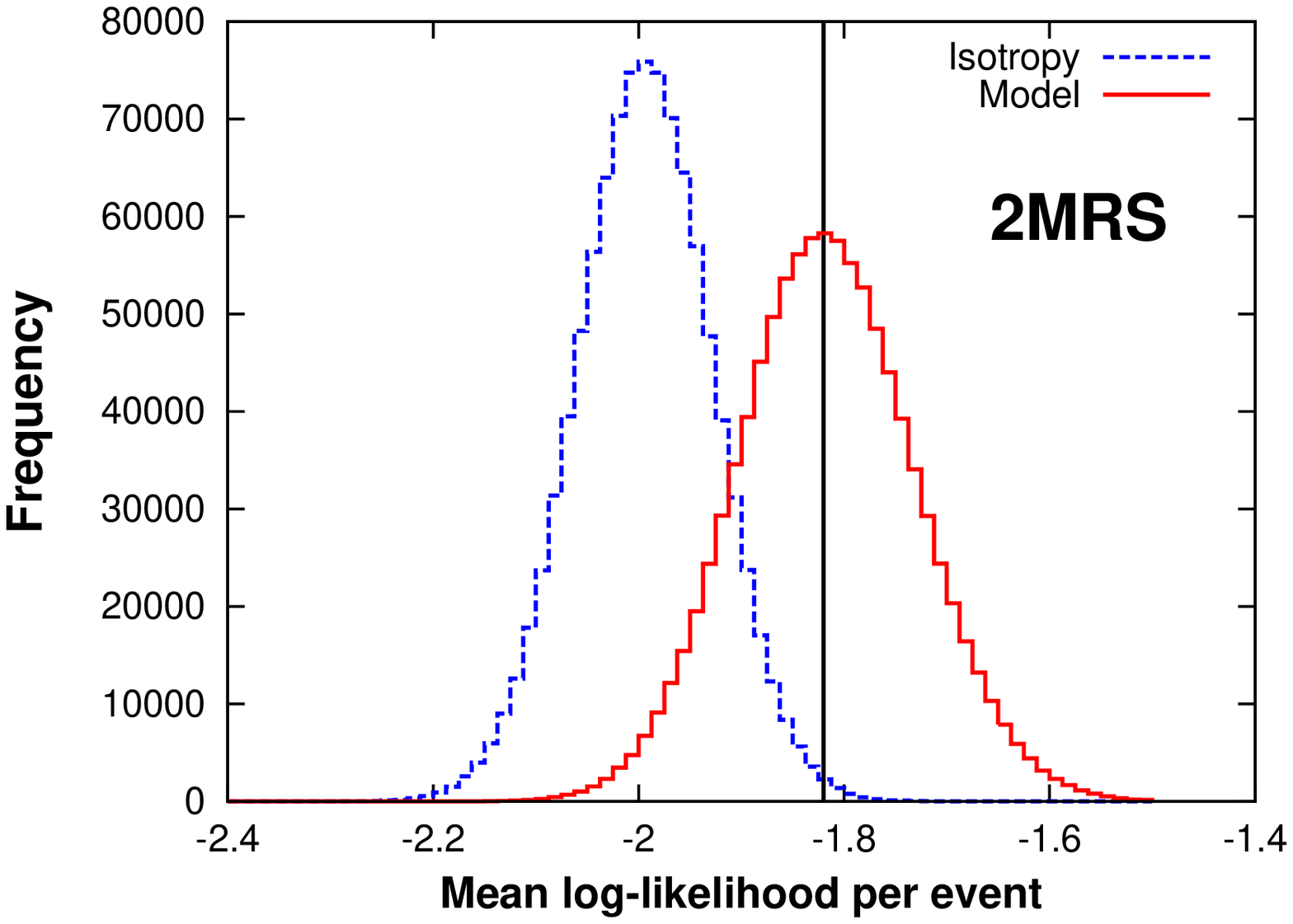}
\hfill\includegraphics[width=0.35\linewidth,angle=-90]{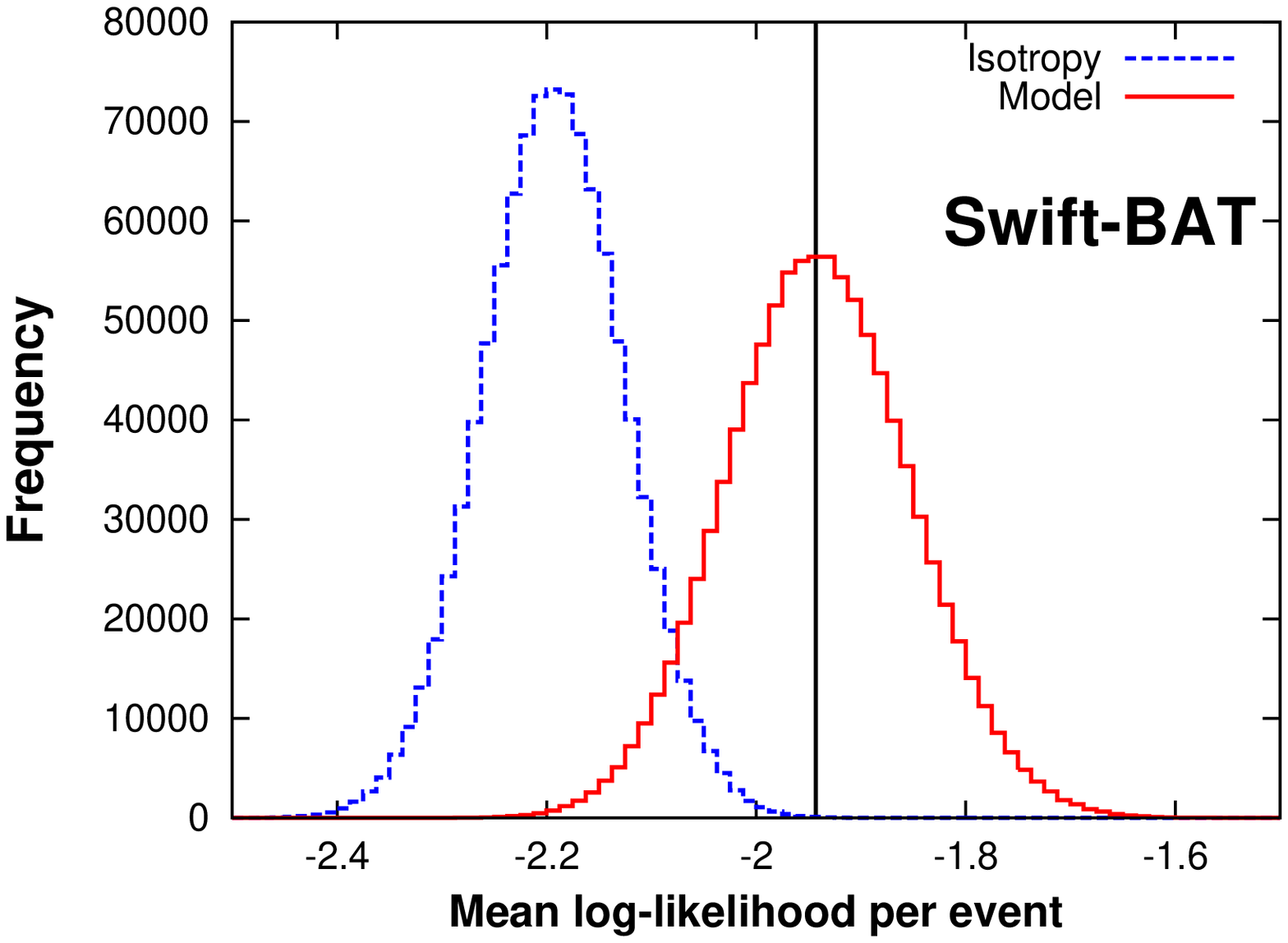}\hfill}
   \centerline{\hfill\includegraphics[width=0.35\linewidth,angle=-90]{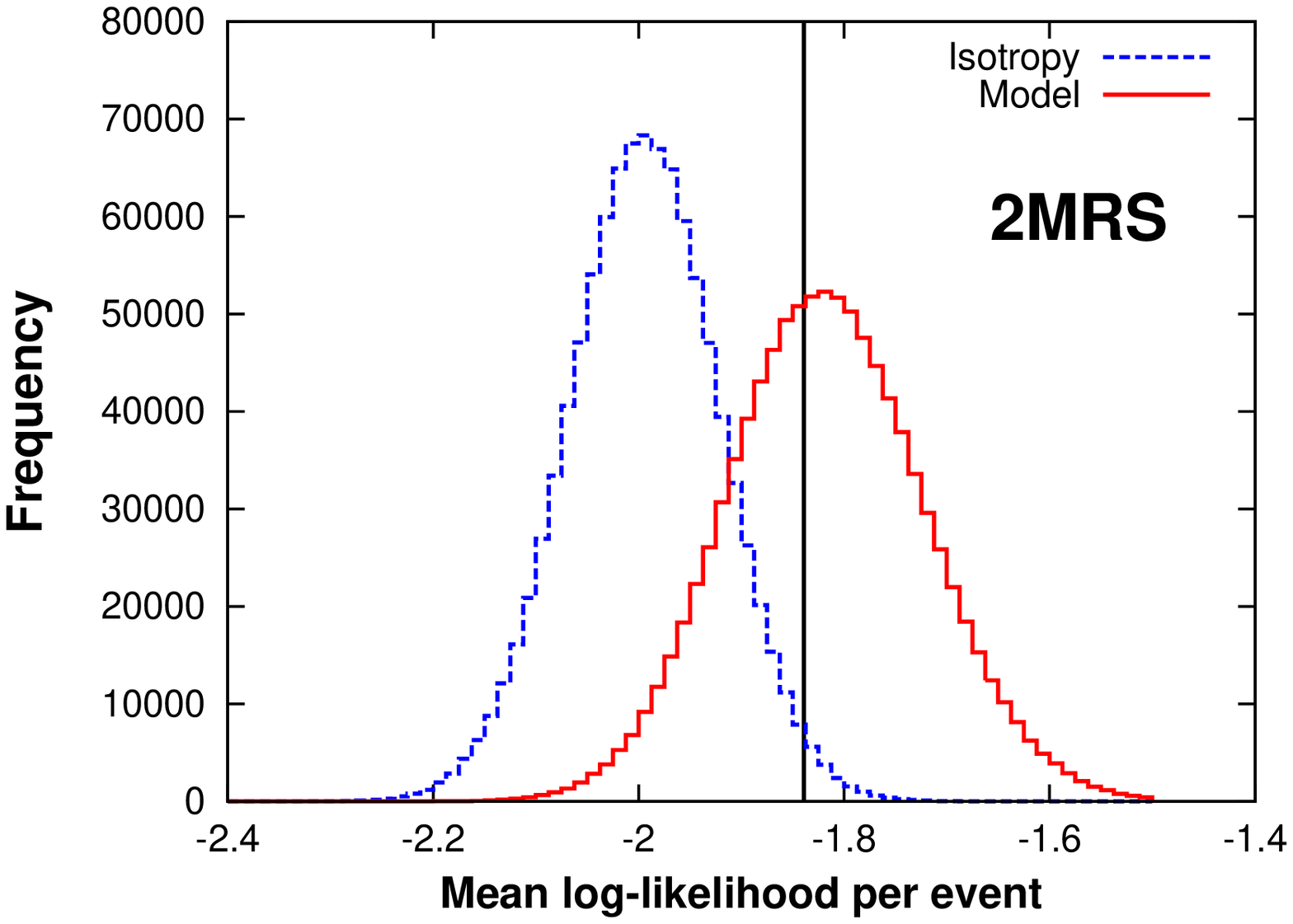}
\hfill\includegraphics[width=0.35\linewidth,angle=-90]{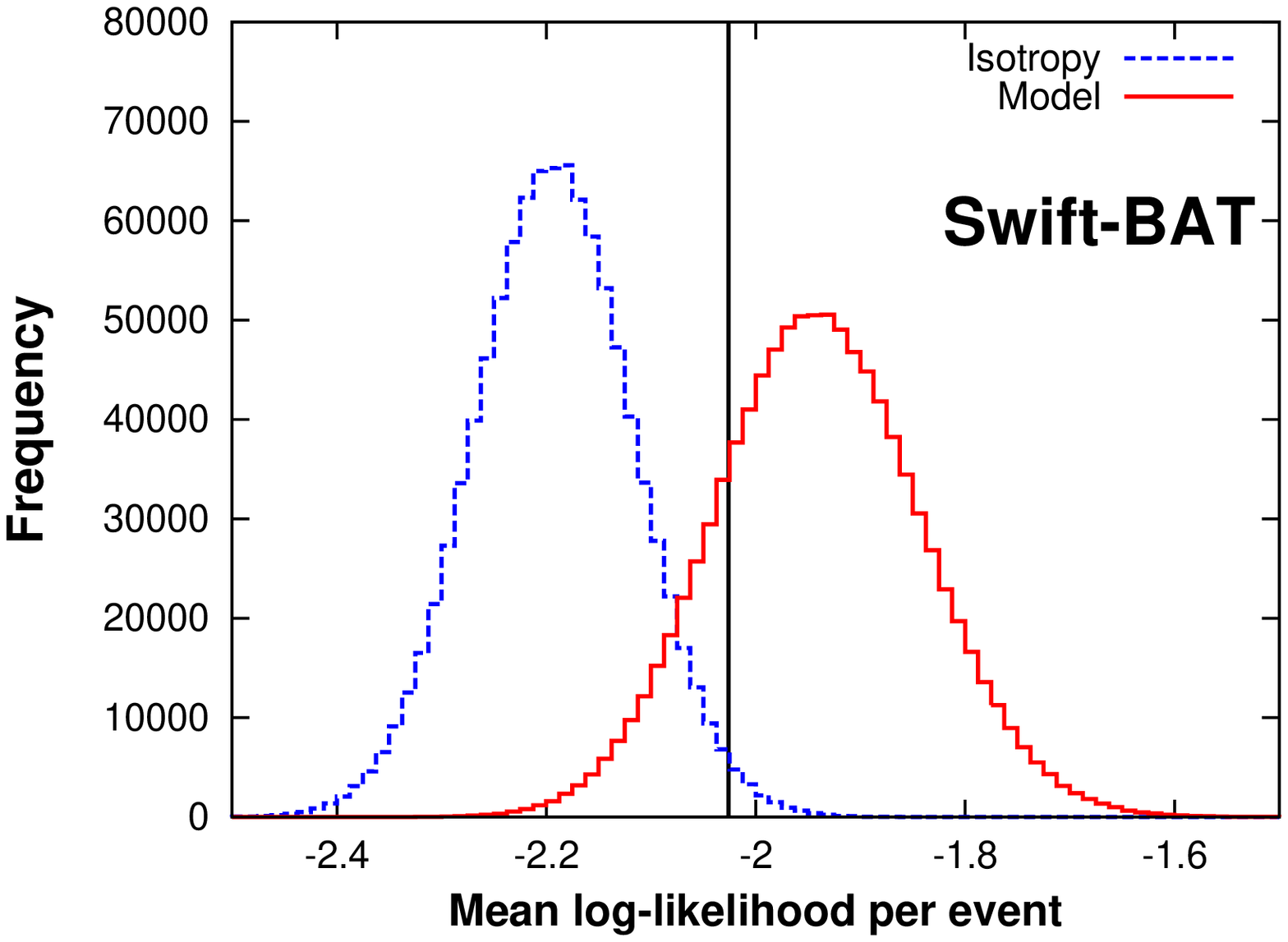}\hfill}
  \caption{Distributions of mean log-likelihood per event for isotropic arrival directions (blue, dashed line histograms) and for the model predictions (red, solid line histograms). The parameters for the models based on the 2MRS galaxies (left) and Swift-BAT AGNs (right) are those that maximize the likelihood with all data, namely $(1.5^\circ, 0.64)$ for 2MRS and $(7.8^\circ,0.56)$ for Swift-BAT. The value of the log-likelihood for the data is indicated by a black vertical line. The plots in the top panels use all data, and those in the bottom panels exclude data collected during period~I.} 
  \label{fig:LL_dist}
 \end{figure}

Finding the values of $\sigma$ and $f_{\rm iso}$ that maximize the log-likelihood does not ensure that the model fits well the data. To test the compatibility between data and model, we generate simulated sets with the same number of arrival directions as in the data, 
drawn either from the density map of the models or isotropically.
We then compare the distributions of the mean log-likelihood ($\mathcal{LL}/N_{\rm data}$) with the value obtained for the data. 
We present the results in Fig.~\ref{fig:LL_dist}. 

Data are compatible with the models and differ 
from average isotropic expectations.  The fraction $f$ of isotropic realizations that have a higher likelihood than the data is  $2\times 10^{-4}$ in the case of the model based on Swift-BAT AGNs, and $4\times 10^{-3}$ with the model based on 2MRS galaxies. These values of $f$ are obtained with the parameters $\sigma$ and $f_{\rm iso}$ that maximize the likelihood for the respective catalogue using all the events with energy larger than 55 EeV (the black dots in the top panels of Fig.~\ref{fig:LL_cont}). With the same parameters, and data from period I excluded,  $f\approx 0.02$ in both models. 
These figures are {\it a posteriori}, and do not represent a confidence level on anisotropy.

The likelihood test is sensitive to whether or not the data points lie
in a high density region of the model. 
Complementary methods can be applied that test the
overall proportionality between the sky distribution of arrival
directions and model predictions. For instance in Ref.~\cite{julien} we have
developed a method based on the smoothed density maps that
simultaneously tests both the correlation as well as the intrinsic
clustering properties of the data compared to the models. 
These tests are inconclusive with present data.
The dispersion in the predictions by different models decreases with an increasing number of events. For instance, the width of the histograms in 
Fig.~\ref{fig:LL_dist} decreases as $1/\sqrt{N}$. With this dispersion reduced by a factor two, if the anisotropy is substantiated by future data it should also become possible to narrow the range of viable astrophysical scenarios.
  
The HiRes collaboration has reported \cite{hiresagn2} 
that their data with threshold energies of 57~EeV are incompatible at 
a 95\% confidence level with a matter tracer model based on 2MRS galaxies 
with smoothing angles smaller than $10^\circ$. The analysis performed in \cite{hiresagn2}
has the smoothing angle as the only free parameter.
As already mentioned at the end of section \ref{vcv}, comparison of results 
between the two observatories is especially challenging around the GZK energy threshold. 
Auger arrival directions are compatible with models of the local matter distribution 
based on 2MRS galaxies for smoothing angles of a few degrees and correlating fractions 
of about 40\% ($f_{\rm iso}\approx 0.6$ is required for the best fit).

\section{Other aspects of the arrival directions\label{other}}

\begin{figure}[H] \centering
\centerline{\includegraphics[angle=-90,width=0.5\linewidth]{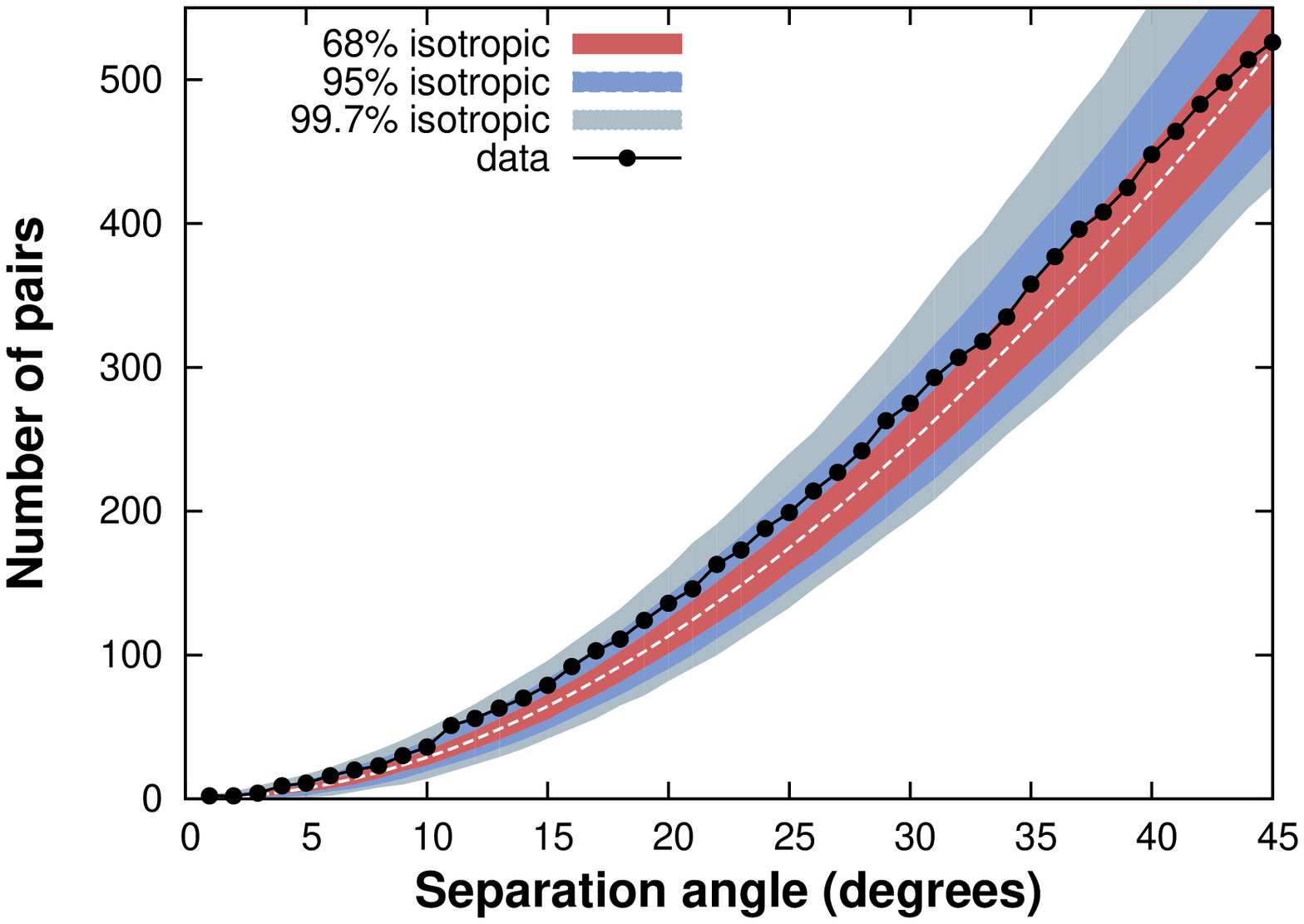}
 \includegraphics[angle=-90,width=0.5\linewidth]{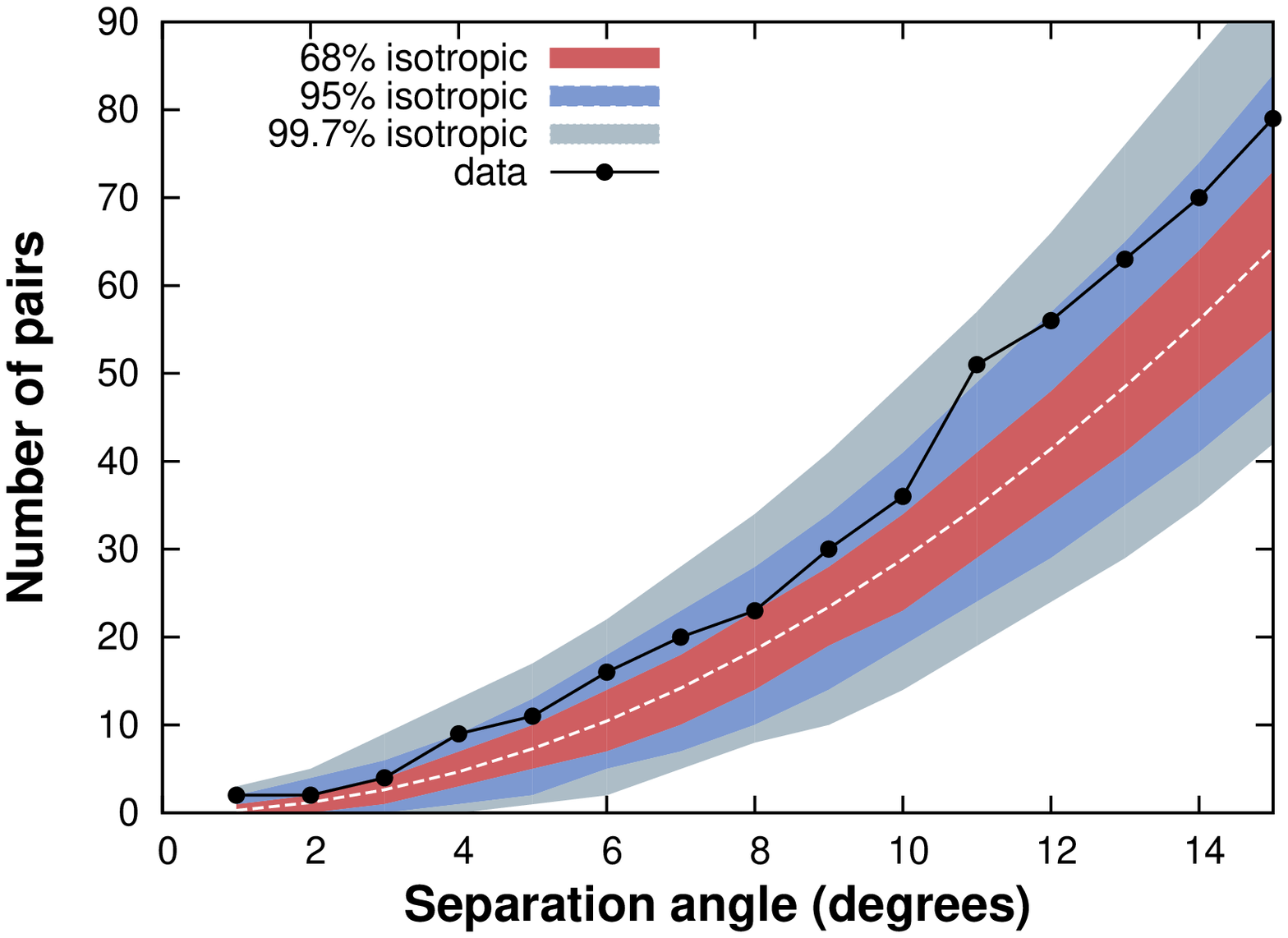}
} 
  \caption{\label{ac} Cumulative autocorrelation function for the set of 69 events
    with $E \ge 55$~EeV (black dots). The bands correspond to the 68\%, 95\% and 99.7\% dispersion 
expected for an isotropic flux. The plot in the
    right panel is an enlarged version of the left plot restricted to
    angles less than $15^\circ$. }
\end{figure}

The autocorrelation of the arrival directions can provide information
about clustering without reference to any catalog.  We show in
Fig.~\ref{ac} the autocorrelation function for the set of the 69
events with $E \ge 55$~EeV. The number of pairs of events with an
angular separation smaller than a given value are plotted as black
dots. The 68\%, 95\%, and 99.7\% dispersion expected in the case of an isotropic flux is
represented by coloured bands. For angles greater than $45^\circ$ (not shown) 
the black dots lie within the 68\%  band.  The region of small
angular scale is shown separately for better resolution.
The largest
deviation from the isotropic expectation occurs for an angular
scale of $11^\circ$, where 51 pairs have a smaller separation compared
with 34.8 pairs expected.  In isotropic realizations of 69 events, a 
fraction $f(11^\circ)=0.013$ have 51 or more pairs within $11^\circ$. 
The fraction of isotropic realizations that achieve $f(\psi)\le 0.013$ 
for any angle $\psi$ is $P=0.10$.

The region with the largest overdensity of arrival directions among
the 69 CRs with $E \ge 55$~EeV, as estimated by the excess above
isotropic expectations in circular windows, is centred at
galactic coordinates ($l,b)=(-46.4^\circ,17.7^\circ$). There are 12
arrival directions inside a window with radius $13^\circ$ centred in
that location, where 1.7 is the isotropic expectation. The
centre of this region is only $4^\circ$ away from the location of the
radiogalaxy Cen A $(-50.5^\circ,19.4^\circ)$ and it is not far from the
direction of the Centaurus cluster $(-57.6^\circ,21.6^\circ)$.  It was
noted in \cite{pao1,pao2} that the arrival directions of two CR
events correlate with the nucleus position of the radiogalaxy Cen A,
while several lie in the vicinity of its radio lobe extension. At only
3.8 Mpc distance, Cen A is the closest AGN.  It is obviously
an interesting region to monitor with additional data. 

\begin{figure}[H]
\begin{center} 
\includegraphics[width=0.75\linewidth]{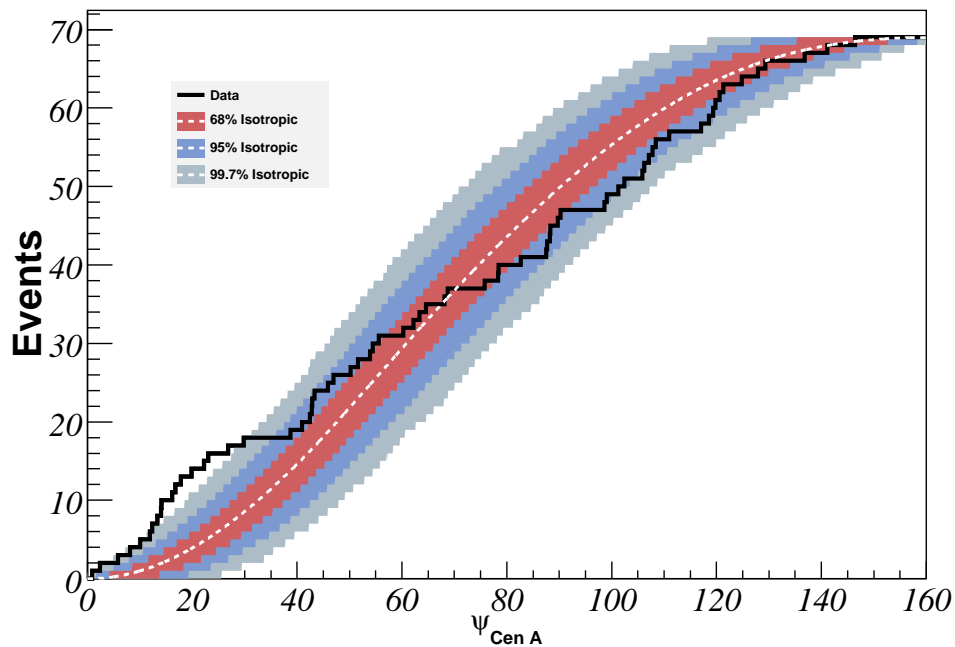} 
\end{center} 
\caption{Cumulative number of events with $E\geq55$ EeV as a function
of angular distance from the direction of Cen A. The bands correspond to the 68\%, 95\%, and 99.7\% dispersion 
expected for an isotropic flux.}\label{cena}
\end{figure}

We show in Fig.~\ref{cena} the number of CR arrival directions within
a variable angular radius from Cen A. In a Kolmogorov-Smirnov test,
4\% of the realizations of 69 arrival directions drawn from an
isotropic distribution have a maximum departure from the 
isotropic expectation greater than or equal to the maximum departure
observed in data. The overdensity with largest significance is given 
by the presence of 13 arrival directions within $18^\circ$, in which 3.2
arrival directions are expected if the flux were isotropic.

The CRs in this region of the sky make a dominant contribution to the
autocorrelation signal. For instance, the 13 arrival directions that
are within $18^\circ$ from Cen A form 6 pairs separated by less than
$4^\circ$, and 28 pairs by less than $11^\circ$. These events also
make a large contribution to the correlation with different
populations of nearby extragalactic objects, both because they are in
excess above isotropic expectations and because this region is densely
populated with galaxies.  The flux-weighted models illustrated in 
Figs.~\ref{swiftweighted} and \ref{probMaps}
predict that the fraction of
CRs inside a circle with radius $18^\circ$ centred at the position of
Cen A is 13.4\% (2MRS) and 29.3\% (Swift-BAT),
compared to 18.8\% observed in data and 4.7\% expected if the flux were isotropic.

In contrast to the region around Cen A and the Centaurus cluster,
there is a paucity of events from the region around the radiogalaxy
M87 and the Virgo cluster. None of the 69 events with $E \ge 55$~EeV is
within $18^\circ$ of M87.  Due to its northern declination, however,
M87 gets only one-third the exposure that Cen A gets at the Southern Auger
observatory.  Only 1.1 events are expected within that $18^\circ$ circle
for an isotropic flux.  

Distance also matters.  M87 is five times farther away than Cen A, so
the flux would be 25 times less if the sources had equal cosmic ray
luminosities.  Coupled with the reduced exposure to M87, the recorded
arrivals from Cen A would be 75 times those from M87 if the two
radiogalaxies were equally luminous in cosmic rays.

The situation is different in comparing the Virgo cluster against the
Centaurus cluster.  While M87 is near the centre of the Virgo cluster,
Cen A is not part of the Centaurus cluster.  Both clusters are well
within the GZK horizon, but the Centaurus cluster is three times more
distant than Virgo.  Combining 1/r$^2$ flux dependence and the
exposure difference, therefore, the recorded events from the Virgo cluster
should outnumber those from Centaurus by three-to-one if the two clusters have
equal cosmic ray luminosities.
The flux-weighted models  illustrated in 
Figs.~\ref{swiftweighted} and \ref{probMaps} 
predict that the fraction of CRs inside a circle with
radius $18^\circ$ centred at the position of M87 is 6.4\% (2MRS)
and 3.0\% (Swift-BAT), compared to 
1.6\% expected if the flux were isotropic.

\section{Discussion and conclusions\label{conclusions}}

Between January 2004 and December 2009 the Pierre Auger Observatory has
detected 69 cosmic rays with energy in excess of 55 EeV. Their arrival
directions are reported here. This data set is more than twice as
large as the one analysed in \cite{pao1,pao2}, which provided evidence
of anisotropy in CR arrival directions at the 99\% confidence
level. The anisotropy was tested with {\it a priori} parameters through the
correlation between the arrival directions of CRs and the positions of
nearby active galaxies from the VCV catalog.  The degree of that
observed correlation has decreased from $(69^{+11}_{-13})\%$ to
$(38^{+7}_{-6})\%$, to be compared with the 21\% expected to
occur by chance if the flux were isotropic.  More data are needed to
determine  this correlating fraction accurately.

We have further examined with {\it a posteriori} explorations the arrival 
directions of these CRs using different scenarios.  
We have compared the distribution of
arrival directions with the positions of different populations of
nearby extragalactic objects: galaxies in the 2MRS survey
and AGNs detected in X-rays by Swift-BAT. We have considered models in
which the CR luminosity is proportional to the flux in the respective
wavelength for the objects in these catalogs. Data are readily
compatible with the models for suitable parameters (smoothing angle
$\sigma$ and isotropic fraction $f_{\rm iso}$). The values of these
parameters have been obtained for each model as best fits to the data:
they are around a few degrees for $\sigma$ and between 0.56 and 0.88
for $f_{\rm iso}$.  Large values of $f_{\rm iso}$ may be an indication of catalog incompleteness, 
or that proportionality between CR luminosity and electromagnetic flux is unrealistic, 
or that a fraction of the arrival directions  
are isotropized by large magnetic deflections due to large charges and/or encounters with strong field regions. 
The best-fit values of $\sigma$ and $f_{\rm iso}$ are not strongly constrained with the present
statistics. These studies are {\it a posteriori} and do not
constitute further quantitative evidence for anisotropy.  They show
that, at present, there are multiple astrophysical models of
anisotropy arising from the distribution of matter in the nearby
universe which are fully consistent with the observed distribution of
arrival directions. 

The autocorrelation of the arrival directions shows only a modest excess of 
direction pairs over a broad range of small angles. In scenarios of discrete sources in
the nearby universe, the absence of strong clustering at small angles
can be interpreted as evidence of many contributing sources and/or
large angular separations between arrival directions from the same
source.  

We have analyzed the region of the sky close to the location of the
radiogalaxy Cen A, since this corresponds to the largest observed
excess with respect to isotropic expectations. The CRs in this region
make a strong contribution to the autocorrelation signal and to the
correlation with different populations of nearby extragalactic
objects. From all the arrival directions of CRs with $E \ge 55$~EeV, 18.8\%
lie within $18^\circ$ of Cen A, while 4.7\% is the isotropic expectation. 
This region is densely populated with
different types of nearby extragalactic objects. Flux-weighted models
based on 2MRS galaxies and on Swift-BAT AGNs  predict a fraction 
of CRs from this region  of 13\% and 29\% respectively.  As reported in 2007
\cite{pao1,pao2}, there are two arrival directions very close to the
position of the Cen A nucleus.  Aside from those two events, the excess is
distributed rather broadly.

Measurements by the Pierre Auger Observatory \cite{paocomposition} of
the depth of shower maximum and its fluctuations indicate a trend
toward heavy nuclei with increasing energy.  Although the measurements
available now are only up to about 55 EeV, the trend suggests that primary CRs
are likely to be dominated by heavy nuclei at higher energies.  This interpretation
of the shower depths is not certain, however.  It relies on shower
simulations that use hadronic interaction models to extrapolate
particle interaction properties two orders of magnitude in
center-of-mass energy beyond the regime where they have been tested
experimentally.  A knowledge of
 CR composition is important for deciding which of several source
scenarios is more likely.  
The trajectories of highly charged nuclei are expected to
undergo large deflections due to the Galaxy's magnetic fields.  While
a correlation of arrival directions with nearby matter on small
angular scales is plausible for protons above 55 EeV, it is puzzling
if the CRs are heavy nuclei.

Definitive conclusions must await additional data. The correlation of 
recent data with objects in the VCV catalog is not as strong as that observed 
in 2007. If the evidence for anisotropy is substantiated by future data, 
then it should also become possible to discriminate between different astrophysical 
scenarios using techniques of the type that have been presented here to explore the 
compatibility of different models with the present set of arrival directions.

\section*{Acknowledgments}
The successful installation and commissioning of the Pierre Auger Observatory
would not have been possible without the strong commitment and effort
from the technical and administrative staff in Malarg\"ue.

We are very grateful to the following agencies and organizations for financial support: 
Comisi\'on Nacional de Energ\'{\i}a At\'omica, 
Fundaci\'on Antorchas,
Gobierno De La Provincia de Mendoza, 
Municipalidad de Malarg\"ue,
NDM Holdings and Valle Las Le\~nas, in gratitude for their continuing
cooperation over land access, Argentina; 
the Australian Research Council;
Conselho Nacional de Desenvolvimento Cient\'{\i}fico e Tecnol\'ogico (CNPq),
Financiadora de Estudos e Projetos (FINEP),
Funda\c{c}\~ao de Amparo \`a Pesquisa do Estado de Rio de Janeiro (FAPERJ),
Funda\c{c}\~ao de Amparo \`a Pesquisa do Estado de S\~ao Paulo (FAPESP),
Minist\'erio de Ci\^{e}ncia e Tecnologia (MCT), Brazil;
AVCR AV0Z10100502 and AV0Z10100522,
GAAV KJB300100801 and KJB100100904,
MSMT-CR LA08016, LC527, 1M06002, and MSM0021620859, Czech Republic;
Centre de Calcul IN2P3/CNRS, 
Centre National de la Recherche Scientifique (CNRS),
Conseil R\'egional Ile-de-France,
D\'epartement  Physique Nucl\'eaire et Corpusculaire (PNC-IN2P3/CNRS),
D\'epartement Sciences de l'Univers (SDU-INSU/CNRS), France;
Bundesministerium f\"ur Bildung und Forschung (BMBF),
Deutsche Forschungsgemeinschaft (DFG),
Finanzministerium Baden-W\"urttemberg,
Helmholtz-Ge\-mein\-schaft Deutscher Forschungszentren (HGF),
Ministerium f\"ur Wissenschaft und Forschung, Nordrhein-Westfalen,
Ministerium f\"ur Wissenschaft, Forschung und Kunst, Baden-W\"urt\-temberg, Germany; 
Istituto Nazionale di Fisica Nucleare (INFN),
Istituto Nazionale di Astrofisica (INAF),
Ministero dell'Istruzione, dell'Universit\`a e della Ricerca (MIUR), Italy;
Consejo Nacional de Ciencia y Tecnolog\'{\i}a (CONACYT), Mexico;
Ministerie van Onderwijs, Cultuur en Wetenschap,
Nederlandse Organisatie voor Wetenschappelijk Onderzoek (NWO),
Stichting voor Fundamenteel Onderzoek der Materie (FOM), Netherlands;
Ministry of Science and Higher Education,
Grant Nos. 1 P03 D 014 30 and N N202 207238, Poland;
Funda\c{c}\~ao para a Ci\^{e}ncia e a Tecnologia, Portugal;
Ministry for Higher Education, Science, and Technology,
Slovenian Research Agency, Slovenia;
Comunidad de Madrid, 
Consejer\'{\i}a de Educaci\'on de la Comunidad de Castilla La Mancha, 
FEDER funds, 
Ministerio de Ciencia e Innovaci\'on and Consolider-Ingenio 2010 (CPAN),
Generalitat Valenciana, 
Junta de Andaluc\'{\i}a, 
Xunta de Galicia, Spain;
Science and Technology Facilities Council, United Kingdom;
Department of Energy, Contract Nos. DE-AC02-07CH11359, DE-FR02-04ER41300,
National Science Foundation, Grant No. 0450696,
The Grainger Foundation USA; 
ALFA-EC / HELEN,
European Union 6th Framework Program,
Grant No. MEIF-CT-2005-025057, 
European Union 7th Framework Program, Grant No. PIEF-GA-2008-220240,
and UNESCO.

\appendix
\section{Event list}
We list in the following table the equatorial coordinates (RA, Dec) and the galactic coordinates ($l$, $b$) of the 69 events recorded from 1 January 2004 up to 31 December 2009 with $E \ge 55$~EeV, together with their date of observation (year and Julian day), zenith angle ($\theta$), signal at 1000~m from the shower core $S(1000)$, and energy $E$. $S(1000)$ is measured in units called VEM, determined by the average charge
deposited by a high-energy down-going vertical
and central muon \cite{paoproperties}. The energy resolution is about 15\% and the absolute
energy scale has a systematic uncertainty of 22\% \cite{paoflux,paoflux2}. The angular resolution is better than
$0.9^\circ$ \cite{angularresolution}.

In \cite{pao2} we published the list of the first 27 events, detected in periods I and II in Table~\ref{periods}. Since then, the reconstruction algorithms and calibration procedures of the Pierre Auger Observatory have been updated and refined. The lowest energy among these same 27 events (which was 57~EeV in \cite{pao2}) is $55$~EeV according to the latest reconstruction. The reconstructed values of $S(1000)$ have changed by less than 4\% and of the energy by less than 7\%. The arrival directions of 26 events differ by less than $0.1^\circ$ from their previous determination, while one differs by $0.4^\circ$. 
Events recorded in periods I, II, and III are separated by horizontal lines.

\begin{center}
\begin{longtable}{ccccccccc}
\hline Year & Julian day & $\theta$ (deg) & S(1000) & $E$ (EeV) & RA (deg) & Dec (deg) & $l$ (deg) & $b$ (deg) \\ \hline
\endfirsthead

\hline
Year & Julian day & $\theta$ (deg) & S(1000) & $E$ (EeV) & RA (deg) & Dec (deg) & $l$ (deg) & $b$ (deg) \\  \hline
\endhead

\hline \multicolumn{9}{|r|}{{Continued on next page}} \\ \hline
\endfoot

\hline \hline
\endlastfoot
			2004 & 125 & 47.7 & 245 & 65 & 267.1  & -11.4 &    15.5 &    8.4  \\
                        2004 & 142 & 59.3 & 205 & 79 & 199.7  & -34.9 &   -50.7 &  27.7  \\
                        2004 & 282 & 26.5 & 329 & 64 & 208.1  & -60.3 &   -49.6 &    1.7  \\
                        2004 & 339 & 44.7 & 324 & 83 & 268.6  & -60.9 &   -27.6 & -17.0 \\
                        2004 & 343 & 23.4 & 321 & 60 & 224.5  & -44.2 &   -34.3 &  13.0  \\
                        2005 &   54 & 35.0 & 374 & 81 &   17.4  & -37.9 &   -75.6 & -78.6 \\
                        2005 &   63 & 54.4 & 214 & 68 & 331.2  &   -1.2 &    58.7 & -42.4  \\
                        2005 &   81 & 17.1 & 309 & 55 & 199.1  & -48.5 &   -52.8 &  14.1  \\
                        2005 & 295 & 15.4 & 310 & 55 & 332.9  & -38.2 &      4.2 & -54.9 \\
                        2005 & 306 & 40.0 & 248 & 56 & 315.4  &   -0.4 &    48.8 & -28.8  \\
                        2005 & 306 & 14.2 & 444 & 80 & 114.6  & -43.0 & -103.8 & -10.3  \\
                        2006 &   35 & 30.8 & 396 & 82 &  53.7  &   -7.8 & -165.9 & -46.9  \\
                        2006 &   55 & 37.9 & 264 & 58 & 267.7  & -60.6 &   -27.5 & -16.5  \\
                        2006 &   81 & 34.0 & 367 & 78 & 201.1  & -55.3 &   -52.3 &    7.3  \\
                        \hline 
                        2006 & 185 & 59.0 & 211 & 80 & 350.1  &    9.5 &    88.7 &  -47.2  \\
                        2006 & 296 & 54.0 & 207 & 66 &   53.0  &   -4.2 & -170.7 &  -45.4  \\
                        2006 & 299 & 26.0 & 344 & 66 & 200.9  & -45.3 &   -51.2 &   17.2  \\
                        2007 &   13 & 14.3 & 753 & 142 & 192.8  & -21.1 &   -57.2 &   41.8  \\
                        2007 &   51 & 39.2 & 255 & 57 & 331.7  &    2.9 &     63.5 & -40.3  \\
                        2007 &   69 & 30.4 & 334 & 68 & 200.2  & -43.3 &   -51.4 &   19.3  \\
                        2007 &   84 & 17.2 & 341 & 61 & 143.2  & -18.3 & -109.3 &   23.8  \\
                        2007 & 145 & 23.9 & 400 & 77 &  47.6  & -12.8 & -164.0 &  -54.5  \\
                        2007 & 186 & 44.8 & 254 & 64 & 219.4  & -53.8 &   -41.7 &     5.8 \\
                        2007 & 193 & 17.9 & 470 & 87 & 325.5  & -33.4 &    12.2 &  -49.0 \\
                        2007 & 221 & 35.3 & 318 & 68 & 212.7  &   -3.2 &   -21.8 &   54.1  \\
                        2007 & 234 & 33.3 & 366 & 77 & 185.3  & -27.9 &   -65.2 &   34.5  \\
                        2007 & 235 & 42.6 & 275 & 66 & 105.9  & -22.9 & -125.2 &    -7.7  \\
                        \hline
			2007 & 295 & 21.1 & 389 & 73 & 325.7  & -15.6 &    37.8 & -44.8 \\
                        2007 & 343 & 30.9 & 447 & 93 &   81.5  &   -7.4 & -150.1 & -22.3 \\
                        2007 & 345 & 51.5 & 212 & 62 & 314.9  & -53.4 &   -15.5 & -40.4 \\
                        2008 &   13 & 17.0 & 363 & 66 & 252.8  & -22.6 &     -1.8 &  13.7 \\
                        2008 &   18 & 50.1 & 389 & 115 & 352.7  & -20.9 &    47.4 & -70.5 \\
                        2008 &   36 & 28.4 & 367 & 73 & 186.9  & -63.6 &   -59.7 &  -0.9 \\
                        2008 &   51 & 20.7 & 314 & 58 & 201.9  & -54.9 &   -51.8 &   7.6 \\
                        2008 &   52 & 31.7 & 308 & 63 &  82.8  & -15.8 & -141.2 & -24.7 \\
                        2008 &   87 & 39.0 & 355 & 82 & 220.5  & -42.9 &   -36.4 &  15.5 \\
                        2008 & 118 & 36.2 & 324 & 70 & 110.2  &   -0.9 & -142.9 &    6.2 \\
                        2008 & 192 & 20.4 & 302 & 55 & 306.7  & -55.3 &   -17.3 & -35.4 \\
                        2008 & 205 & 53.0 & 183 & 56 & 358.9  &  15.5 &  103.6 & -45.3 \\
                        2008 & 264 & 44.4 & 384 & 99 & 116.0  & -50.6 &   -96.4 & -12.9 \\
                        2008 & 268 & 49.8 & 415 & 123 & 287.6  &    1.5 &     36.4 &  -3.6 \\
                        2008 & 282 & 28.9 & 309 & 61 & 202.3  & -16.1 &   -44.0 &  45.8 \\
                        2008 & 296 & 42.8 & 293 & 71 &  15.6  & -17.0 &  137.7 & -79.6 \\
                        2008 & 322 & 28.3 & 345 & 68 &  25.1  & -61.2 &   -67.3 & -54.9 \\
                        2008 & 328 & 47.2 & 250 & 66 & 126.5  &    5.3 & -140.8 &  23.4 \\
                        2008 & 337 & 31.0 & 348 & 71 & 275.5  & -14.4 &    16.8 &   -0.1 \\
                        2008 & 362 & 31.4 & 406 & 84 & 209.6  & -31.3 &   -40.7 &  29.4 \\
                        2009 &     7 & 59.3 & 152 & 57 & 286.3  & -37.6 &     -0.5 & -18.7 \\
                        2009 &   30 & 32.3 & 346 & 72 & 303.9  & -16.7 &    26.6 & -25.9 \\
                        2009 &   32 & 56.2 & 199 & 67 &     0.0  & -15.4 &    75.0 &  -73.3\\
                        2009 &   35 & 52.8 & 191 & 57 & 227.1  & -85.2 &   -54.1 & -23.2 \\
                        2009 &   39 & 42.4 & 291 & 70 & 147.2  & -18.3 & -106.5 &  26.6 \\
                        2009 &   47 & 20.8 & 311 & 57 &   78.3  & -16.0 & -142.9 & -28.8 \\
                        2009 &   51 &   7.1 & 377 & 65 & 203.7  & -33.1 &   -46.7 &  28.9 \\
                        2009 &   78 &   8.2 & 350 & 61 &  26.7  & -29.1 & -134.6 & -77.6 \\
                        2009 &   78 & 27.3 & 424 & 84 & 122.9  & -54.6 &   -90.7 & -11.3 \\
                        2009 &   80 & 44.5 & 263 & 66 & 170.1  & -27.1 &   -80.9 &  31.5 \\
                        2009 &   80 & 18.4 & 388 & 71 &  251.4  & -35.8 &   -13.0 &    6.3 \\
                        2009 & 160 & 40.9 & 242 & 56 &    43.8  & -25.5 & -143.2 & -62.3 \\
                        2009 & 168 & 27.0 & 294 & 57 &  153.6 &  -8.6 & -109.4 &  37.9 \\
                        2009 & 191 & 26.9 & 339 & 66 &  294.5 & -20.5 & 19.1 & -19.2 \\
                        2009 & 212 & 52.7 & 188 & 57 & 122.6  & -78.5 & -68.8 & -22.8 \\
                        2009 & 219 & 40.2 & 252 & 57 & 29.4 & -8.6 & 166.1 & -65.8 \\
                        2009 & 225 & 26.2 & 298 & 57 & 90.5 & -21.3 & -132.8 & -20.0 \\
                        2009 & 262 & 22.4 & 341 & 64 & 50.1 & -25.9 & -140.5 & -56.7 \\
                        2009 & 282 & 47.2 & 231 & 61 & 47.7 & 11.5 & 168.7 & -38.6 \\
                        2009 & 288 & 34.2 & 310 & 66 & 217.9 & -51.5 & -41.6 & 8.3 \\
                        2009 & 304 & 30.1 & 304 & 61 & 177.7 & -5.0 & -83.8 & 54.7 \\
                        2009 & 326 & 31.4 & 283 & 57 & 5.4 & -5.6 & 103.3 & -67.3 \\
\end{longtable}
\end{center}

\end{document}